\documentclass{article}
\usepackage{arxiv}
\usepackage[utf8]{inputenc} 
\usepackage{hyperref}       
\usepackage{braket}
\usepackage{amsmath}
\usepackage{amssymb} 
\usepackage{graphicx}
\usepackage{mathrsfs}
\usepackage{cite}
\usepackage{subfigure}
\title{Graphene generalized coherent states}
\author{David J. Fern{\'a}ndez C. and Daniel O-Campa\\
Physics Department, Cinvestav\\
P.O.B. 14-740, 07000 Mexico City, Mexico\\
e-mail: david@fis.cinvestav.mx, dortiz@fis.cinvestav.mx
}
\begin{document}
\maketitle
\begin{abstract}
In this paper we construct the generalized coherent states for an electron in monolayer or bilayer graphene placed in an external magnetic field. At first we define an appropriate set of ladder operators acting on the eigenfunctions for each Hamiltonian, afterwards we derive the generalized coherent states through several definitions and analyze the commutation relationship between the ladder operators. Then we determine the conditions leading to the mutual equivalence between coherent states. Finally, some physical quantities are calculated and we study the time evolution for the generalized graphene coherent states. 
\end{abstract}
\section{Introduccion}
The subject coherent states is tightly related to the harmonic oscillator. In fact, the coherent states were originally found by Schrödinger as minimal uncertainty states for the harmonic oscillator \cite{Schrodinger1926}, but it was not until the sixties of the previous century that Glauber, trying to describe the electromagnetic field coherence, coined the term \cite{PhysRev.131.2766}. The harmonic oscillator coherent states can be achieved by three mutually equivalent definitions: 1) they are eigenstates of the annihilation operator, called Barut–Girardello coherent states in the literature \cite{Barut:1970qf}; 2) they are states resulting from the action of the displacement operator onto the ground state, also known as Gilmore-Perelomov coherent states \cite{cmp/1103858078,gilmore2012lie}; 3) they are states with the minimal uncertainty \cite{Schrodinger1926,PhysRevLett.41.207,PhysRevD.20.1321}. Each of these definitions can be employed to get coherent states for arbitrary systems, but in general the mutual equivalence between these coherent states will be lost (see e.g. \cite{C_1994,C_1999,BERMUDEZ2014615,ContrerasAstorga2022}).

On the other hand, the electrons behavior in graphene has been extensively studied due to the interesting physical 
properties of these materials. Usually this electronic structure is determined by using the tight-binding model\cite{RevModPhys.81.109,katsnelson_2012,McCann_2013}. Different authors have focused on finding exact solutions for the stationary states of electrons in graphene placed in external electromagnetic fields \cite{Kuru_2009,Fern_ndez_C_2020,Midya_2014,Fernandez_2021,CASTILLOCELEITA2020168287}. Recently, the use of coherent states has gained relevance since they have properties different from the stationary states \cite{Diaz-Bautista2017,D_az_Bautista_2019,Fernandez2020,Diaz-Bautista2021,D_az_Bautista_2020}. However, in each of these works to obtain these states depends on the coherent state definition used.

In this work we intend to find these states by studying the three previously mentioned definitions. Furthermore, a path allowing us to get the mutual equivalence between these states is explored. For doing this, the paper has been organized as follows: in section 2 a brief description of the effective Hamiltonian for monolayer and bilayer graphene is presented, as well as some details leading us to exact solutions for the stationary states. In section 3, the corresponding ladder operators for monolayer and bilayer graphene respectively are discussed. In section 4, the coherent states are constructed by means of three different definitions, the commutation relation between the ladder operators is analyzed and the conditions leading us to the mutual equivalence between these definitions are sought. In section 5, the generalized coherent states for a constant magnetic field are found and some physical properties are explored. Finally, in section 6 the conclusions of this paper are established.
\section{Effective Hamiltonians of graphene}
The tight-binding model allows to work with an effective Hamiltonian which rules the electron motion in monolayer or bilayer graphene. If magnetic fields are involved, these Hamiltonians are modified according to the minimal coupling rule \cite{katsnelson_2012,Kuru_2009}. Assuming that the applied magnetic fields are orthogonal to monolayer or the bilayer graphene surface ($x-y$ plane), different gauges can be chosen. We will use the Landau gauge and we will assume that the field changes only along a fixed direction (namely $x$). Thus, the vector potential  is chosen as $\vec{A}=\mathcal{A}(x)\hat{e}_y$ and then $\vec{B}=\mathfrak{B}(x)\hat{e}_z$, with $\mathfrak{B}(x)=\mathcal{A}'(x)$. Next we will discuss the eigenvalue problem for each graphene configuration.
\subsection{Monolayer graphene}
The effective Hamiltonian for an electron in monolayer graphene at low energies after applying the minimal coupling rule is given by 
\begin{equation}
\mathcal{H}_{\mathcal{M}}=v_{{\scriptscriptstyle F}}\vec{\sigma}\cdot\vec{\pi},
\label{1}
\end{equation}
with $\vec{\pi}=\vec{p}+(e/c)\vec{A}$ being known as the kinematical momentum, $v_{{\scriptscriptstyle F}}$ is the Fermi velocity, $e$ the electron charge, $c$ the speed of light and $\vec{\sigma}=(\sigma_1,\sigma_2)$ is an array which components are the Pauli matrices. For stationary states the time-independent eigenvalue equation reads
\begin{equation}
\mathcal{H}_{\mathcal{M}}\Psi(x,y)
=\left(v_{{\scriptscriptstyle F}}\vec{\sigma}\cdot\vec{\pi}\right)\Psi(x,y)=E\Psi(x,y).
\label{2}
\end{equation}
Due to the assumptions about the magnetic field made previously, $\mathcal{H}_{\mathcal{M}}$ and $p_y$ commute, hence the eigenfunctions can be written as
\begin{equation}
\Psi\left(x,y\right)=e^{iky}
\begin{pmatrix} 
\psi^+\left(x\right)\\ 
i\psi^-\left(x\right)
\end{pmatrix}, 
\label{3}
\end{equation}
and thus equations (\ref{2}) and (\ref{3}) lead us to the following system of coupled first-order differential equations for
$\psi^{\pm}$:
\begin{equation}
\mathcal{L}_1^{\pm}\psi^{\pm}(x)=\sqrt{\mathcal{E}}\psi^{\mp}(x),
\label{4}
\end{equation}
where
\begin{equation}
\mathcal{L}_1^{\pm}=\mp\frac{d}{dx}+W(x),
\qquad
W(x)=\frac{e}{c\hbar}\mathcal{A}(x)+k,
\qquad
\mathcal{E}=\frac{E^2}{\hbar^2v^2_{{\scriptscriptstyle F}}}.
\label{5}
\end{equation}
The system of equations (\ref{4}) can be decoupled, leading to
\begin{equation}
H_1^\pm\psi^\pm=\mathcal{E}\psi^{\pm}.
\label{6}
\end{equation}
The operators $H_1^{\pm}$ have the form of two Schrödinger Hamiltonians, 
\begin{equation}
H_1^\pm=\mathcal{L}_1^\mp \mathcal{L}_1^\pm=-\frac{d^2}{dx^2}+V^{\pm}(x),
\qquad
\qquad
V^{\pm}(x)=W^{2}(x)\pm W'(x),
\label{7}
\end{equation}
which fulfill the following intertwining relationships
\begin{equation}
H_1^{\pm}\mathcal{L}_1^{\mp}=\mathcal{L}_1^{\mp}H_1^{\mp}.
\label{8}
\end{equation}
In the context of supersymmetric quantum mechanics (SUSY QM) the operators $H_1^{\pm}$ ($V^{\pm}$) are known as partner Hamiltonians (partner potentials), $W$ is called the superpotential and $\mathcal{L}_1^{\pm}$ are the intertwining operators.

Equation (\ref{8}) gives us the chance to solve the eigenvalue problem (\ref{2}): we just need to solve one of the two uncoupled systems (\ref{6}), the other one automatically follows from it. Thus, let us suppose that we know the eigenfunctions $\psi_n^-$ and eigenvalues $\mathcal{E}_n$ of $H_1^-$ and that the ground state is annihilated by $\mathcal{L}_1^{-}$, the corresponding eigenvalue being $\mathcal{E}_0=0$ (see \cite{Kuru_2009} and references therein). Then, the eigenfunctions of $H_1^+$ read as follows
\begin{equation}
\psi^{+}_n(x)=\frac{1}{\sqrt{\mathcal{E}_n}}\mathcal{L}_1^{-}\psi^{-}_n(x),
\qquad
\mbox{for}
\qquad
n=1,2, ...
\label{9}
\end{equation}
with eigenvalues $\mathcal{E}_n$. Finally, the eigenfunctions and eigenvalues of equation (\ref{2}) are given by
\begin{align}
\Psi_0(x,y)&=c_0e^{iky}
\begin{pmatrix} 
0\\ 
i\psi_0^-(x)
\end{pmatrix},
\qquad
E_0=0,\nonumber\\
\Psi_n(x,y)&=c_ne^{iky}
\begin{pmatrix} 
\frac{1}{\sqrt{\mathcal{E}_n}}\mathcal{L}_1^{-}\psi_n^-(x)\\ 
i\psi_n^-(x)
\end{pmatrix},
\qquad
E_n=\pm\hbar v_{{\scriptscriptstyle F}}\sqrt{\mathcal{E}_n},
\label{10}
\end{align}
where $c_n$ is a normalization constant taking the values $1$ if the spinor top entry is null and $1/\sqrt{2}$ if it is not; the positive energies describe electrons and the negative ones to holes. 

Some important quantities that we will calculate for this system are the probability and current densities, which for an arbitrary state $\Psi$ turn out to be
\begin{equation}
\rho_{\mathcal{M}}=\Psi^{\dagger}(x,y)\Psi(x,y),\qquad\qquad\vec{\mathcal{J}}_{\mathcal{M}}=v_{{\scriptscriptstyle F}}\Psi^{\dagger}(x,y)\vec{\sigma}\Psi(x,y).
\label{11}
\end{equation} 
\subsection{Bilayer graphene}
For bilayer graphene, in addition to consider a magnetic field like the one described above, we will include an interaction term proportional to $\sigma_1$ which is only $x$-dependent, so the effective Hamiltonian becomes
\begin{equation}
\mathcal{H}_{\mathcal{B}}=\frac{1}{2m^{*}}
\begin{pmatrix} 
0 &\left(\pi_x-i\pi_y\right)^{2}
-\hbar^2\beta(x)\\
\left(\pi_x+i\pi_y\right)^{2}
-\hbar^2\beta(x) & 0
\end{pmatrix},
\label{2.2.1}
\end{equation}
where $m^*$ is the electron effective mass and $\beta$ is a real function. The time-independent equation for the stationary states reads
\begin{equation}
\mathcal{H}_{\mathcal{B}}\Psi(x,y)=E\Psi(x,y).
\label{2.2.2}
\end{equation}
Once again, it turns out that $\mathcal{H}_\mathcal{B}$ and $p_y$ commute, thus we can propose the eigenfunctions as
\begin{equation}
\Psi\left(x,y\right)=e^{iky}
\begin{pmatrix} 
\psi^+\left(x\right)\\ 
\psi^-\left(x\right)
\end{pmatrix}.
\label{2.2.3}
\end{equation}
Now, we define $\eta$ and $\beta$ as follows, 
\begin{align}
\eta(x)&=2\left(k+\frac{e}{c\hbar}\mathcal{A}(x)\right),\nonumber\\
\beta(x)&=\frac{\eta^{'}(x)^2-2\eta(x)\eta^{''}(x)-\left(\epsilon_1-\epsilon_2\right)^2}{4\eta^2(x)},
\label{2.2.4}
\end{align}
with $\epsilon_1$, $\epsilon_2$ being two arbitrary constants (in general complex), then the eigenvalue problem (\ref{2.2.2}) is equivalent to
\begin{equation}
\mathcal{L}_2^{\pm}\psi^{\pm}(x)=-\sqrt{\mathscr{E}}\psi^{\mp}(x),
\label{2.2.5}
\end{equation}
where
\begin{equation}
\mathcal{L}_2^{-}=\frac{d^2}{dx^2}+\eta(x)\frac{d}{dx}+\gamma(x),\qquad\mathcal{L}_2^{+}=\frac{d^2}{dx^2}-\eta(x)\frac{d}{dx}+\gamma(x)-\eta'(x),
\qquad
\mathscr{E}=\left(\frac{2m^*E}{\hbar^2}\right)^2
\label{2.2.6}
\end{equation}
and 
\begin{equation}
\gamma(x)=\frac{\eta^2(x)}{4}+\frac{\eta'(x)}{2}-\frac{\eta''(x)}{2\eta(x)}+\bigg(\frac{\eta'(x)}{2\eta(x)}\bigg)^2-\bigg(\frac{\epsilon_1-\epsilon_2}{2\eta(x)}\bigg)^2.
\label{2.2.7}
\end{equation}
Notice that the system of equations (\ref{2.2.5}) can be decoupled in the following way
\begin{align}
\mathcal{L}_2^{+}\mathcal{L}_2^{-}\psi^{-}(x)&=\mathscr{E}\psi^{-}(x), \nonumber\\
\mathcal{L}_2^{-}\mathcal{L}_2^{+}\psi^{+}(x)&=\mathscr{E}\psi^{+}(x).
\label{2.2.8}
\end{align}
In order to solve the eigenvalue problems (\ref{2.2.8}) let us suppose that $\psi^\pm_n(x)$ are the eigenfunctions of two Schrödinger Hamiltonians
\begin{equation}
H_2^{\pm}=-\frac{d^2}{dx^2}+V^{\pm}(x),
\label{2.2.9}
\end{equation}
which are intertwined to each other, namely,
\begin{equation}
H_2^{\pm}\mathcal{L}_2^{\mp}=\mathcal{L}_2^{\mp}H_2^{\mp}.
\label{2.2.10}
\end{equation}  
Using equations (\ref{2.2.6}), (\ref{2.2.9}) and (\ref{2.2.10}) it turns out that
\begin{align}
V^{-}(x)&=-\gamma(x)+\frac{\eta^2(x)}{2}-\frac{\eta'(x)}{2}+\frac{\epsilon_1+\epsilon_2}{2},
\label{2.2.11}\\
V^{+}(x)&=V^{-}(x)+2\eta'(x).
\label{2.2.12}
\end{align}
In the context of SUSY QM $\epsilon_1$ and $\epsilon_2$ are called factorization energies, whose combination of values leads to different cases which have been analyzed elsewhere (see \cite{Fern_ndez_C_2020,Fernandez_2021} and references therein). Here we will focus in the particular case where $\epsilon_1$, $\epsilon_2$ are real and different, related to two seed solutions $u_m^{-}(x),\ m=1,2$ belonging to the kernel of $\mathcal{L}_2^{-}$ and simultaneously being formal eigenfunctions of $H_2^{-}$, $\mathcal{L}_2^{-}u_m^{-}=0$, $H_2^{-}u_m^{-}=\epsilon_m u_m^{-}$. In particular we are going to choose $\epsilon_1=\mathcal{E}_0$, $\epsilon_2=\mathcal{E}_1$, with the seed solutions taken as the corresponding bound state wave-functions.

Then, given the eigenfunctions $\psi_n^{-}$ and eigenvalues $\mathcal{E}_n$  of $H_2^{-}$, the corresponding eigenfunctions and eigenvalues of $H_2^{+}$ will be given by
\begin{equation}
\psi_n^{+}(x)=\frac{\mathcal{L}_2^{-}\psi_n^{-}(x)}{\sqrt{(\mathcal{E}_n-\epsilon_1)(\mathcal{E}_n-\epsilon_{2})}},\qquad\mathcal{E}_n,\qquad\qquad
n=2,3,...
\label{2.2.13}
\end{equation} 

Finally, the eigenfunctions and eigenvalues for the Hamiltonian (\ref{2.2.2}) take the form
\begin{align}
\Psi_n(x,y)&=c_ne^{iky}
\begin{pmatrix} 
0\\ 
\psi_n^-
\end{pmatrix},
\qquad
E_n=0,
\qquad
n=0,1
\nonumber\\
\Psi_n(x,y)&=c_ne^{iky}
\begin{pmatrix} 
\frac{\mathcal{L}_2^{-}\psi_n^{-}(x)}{\sqrt{(\mathcal{E}_n-\mathcal{E}_0)(\mathcal{E}_n-\mathcal{E}_1)}}\\ 
\psi_n^-
\end{pmatrix},
\qquad
E_n=\pm\frac{\hbar^2}{2m^*}\sqrt{(\mathcal{E}_n-\mathcal{E}_0)(\mathcal{E}_n-\mathcal{E}_1)},
\qquad
n\geq2,
\label{2.2.14}
\end{align} 
where $c_n$ is a normalization constant taking the values $1$ if the top entry is null and $1/\sqrt{2}$ if it is not; the positive energies describe electrons and the negative ones to holes.

The probability and current densities to be studied now are given by
\begin{equation}
\rho_{\mathcal{B}}=\Psi^{\dagger}(x,y)\Psi(x,y)\qquad\vec{\mathcal{J}}_{\mathcal{B}}=\frac{\hbar}{m^*}
\left[\mathfrak{Im}\left(\Psi^{\dagger}(x,y)\vec{j}\Psi(x,y)\right)\right],
\label{2.2.15}
\end{equation}
where $\vec{j}$ is an antihermitian operator given by
\begin{align}
j_{x}&=\frac{i}{\hbar}\vec{\sigma}\cdot\vec{\pi},\nonumber\\
j_{y}&=\frac{i}{\hbar}\vec{\chi}\cdot\vec{\pi},
\label{2.2.16}
\end{align}
with $\vec{\chi}=(\sigma_2,-\sigma_1)$ regardless the factorization energies or the gauge chosen. The previous expressions 
agree with the results obtained in \cite{Fernandez_2021} for electrons in bilayer graphene under external magnetic fields; they allow us to recover the right expressions in the absence of magnetic fields, since in that case $\vec{A}=0$ and the canonical and kinematic momentum become equal\cite{PhysRevB.83.165402}.
\section{Ladder operators}
We are interested in constructing the coherent states for each of the two graphene systems previously discussed. In order to do that it is necessary to determine appropriate  ladder operators in both cases. Let us start by assuming that we know proper annihilation and creation operators $\left\lbrace \theta^{-}, \theta^{+}\right\rbrace$ for the $\psi^{-}_n$ of equations (\ref{10}, \ref{2.2.14}), such that
\begin{align}
\theta^{-}\psi_n^{-}&=\sqrt{p_n}\psi_{n-1}^{-},\nonumber\\
\theta^{+}\psi_n^{-}&=\sqrt{q_n}\psi_{n+1}^{-},
\label{3.1}
\end{align}
where $\theta^{+}\equiv(\theta^{-})^{\dagger}$ and the generalized number operator is defined by $N\equiv\theta^{+}\theta^{-}$. We also assume that these operators supply a good ladder relationship, i.e., the ground state is the only one annihilated by $\theta^{-}$ and $q_n\neq 0$ $\forall n$. Note that, in general $\theta^{-}, \theta^{+}$ are not ladder operators for $\psi^+_n$. Nevertheless, in both graphene cases we know the corresponding intertwining operators $\mathcal{L}_1^{\pm}$ and $\mathcal{L}_2^{\pm}$, which is all we need to determine such ladder operators.
\subsection{Ladder operators for monolayer graphene}
Let us start by introducing an annihilation operator for the eigenfunctions (\ref{10}) for monolayer graphene in a way similar to the proposal made in \cite{CASTILLOCELEITA2020168287,Diaz-Bautista2017,D_az_Bautista_2019,Fernandez2020,Diaz-Bautista2021} as follows
\begin{equation}
A_{\mathcal{M}}^{-}=
\begin{pmatrix} 
\mathcal{L}_{1}^{-}\frac{1}{\sqrt{H_{1}^{-}}}\theta^{-}\frac{f_1(H_{1}^{-})}{\sqrt{H_{1}^{-}}}\mathcal{L}_{1}^{+} & -i\mathcal{L}_{1}^{-}\frac{1}{\sqrt{H_{1}^{-}}}\theta^{-}f_1(H_{1}^{-})\\ 
&\\
i\theta^{-}\frac{f_1(H_{1}^{-})}{\sqrt{H_{1}^{-}}}\mathcal{L}_{1}^{+} & \theta^{-}f_1(H_{1}^{-})
\end{pmatrix},
\label{3.1.1}
\end{equation}
where $f_1$ is an auxiliary real function that we can choose at will. Then, the action of $A_{\mathcal{M}}^{-}$ turns out to be:
\begin{equation}
A_{\mathcal{M}}^{-}\Psi_n(x,y)=2\sqrt{p_n}f_1(\mathcal{E}_n)\Psi_{n-1}(x,y)\times
\left\lbrace
\begin{array}{c}
0\quad\mbox{for}\quad n=0,\\
\\
\frac{1}{\sqrt{2}}\quad\mbox{for}\quad n=1,\\
\\
1\quad\mbox{for}\quad n\geq2.
\end{array}
\right.
\label{3.1.2}
\end{equation}
The associated creation operator $A_{\mathcal{M}}^+\equiv\left(A_{\mathcal{M}}^-\right)^{\dagger}$ is given by
\begin{equation}
A_{\mathcal{M}}^{+}=
\begin{pmatrix} 
\mathcal{L}_1^{-}\frac{f_1(H_1^{-})}{\sqrt{H_1^{-}}}\theta^{+}\frac{1}{\sqrt{H_1^{-}}}\mathcal{L}_1^{+} & -i \mathcal{L}_1^{-} \frac{f_1(H_1^{-})}{\sqrt{H_1^{-}}}\theta^{+}\\ 
&\\
if_1(H_1^{-})\theta^{+}\frac{1}{\sqrt{H_1^{-}}}\mathcal{L}_1^{+}& f_1(H_1^{-})\theta^{+}
\end{pmatrix},
\label{3.1.3}
\end{equation}
and its action becomes
\begin{equation}
A_{\mathcal{M}}^{+}\Psi_n(x,y)=2\sqrt{q_n}f_1(\mathcal{E}_{n+1})\Psi_{n+1}(x,y)\times
\left\lbrace
\begin{array}{c}
\frac{1}{\sqrt{2}}\quad\mbox{for}\quad n=0,\\
\\
1\quad\mbox{for}\quad n\geq1.
\end{array}
\right.
\label{3.1.4}
\end{equation}
This equation allows us to construct the $n$-th eigenstate through the successive application of $A_{\mathcal{M}}^+$ over $\Psi_0$, as long as $f_1(\mathcal{E}_{m})\neq 0$ for $m\in[1,n]$.
\subsection{Ladder operators for bilayer graphene}
For this case the annihilation operator is given by
\begin{equation}
A_{\mathcal{B}}^{-}=
\begin{pmatrix} 
\mathcal{L}_2^{-}\frac{1}{\sqrt{(H_2^--\mathcal{E}_0)(H_2^--\mathcal{E}_1)}}\theta^{-}\frac{f_2(H_2^-)}{\sqrt{(H_2^--\mathcal{E}_0)(H_2^--\mathcal{E}_1)}}\mathcal{L}_2^{+} & \mathcal{L}_2^{-}\frac{1}{\sqrt{(H_2^--\mathcal{E}_0)(H_2^--\mathcal{E}_1)}}\theta^{-}f_2(H_2^-)\\ 
&\\
\theta^{-}\frac{f_2(H_2^-)}{\sqrt{(H_2^--\mathcal{E}_0)(H_2^--\mathcal{E}_1)}}\mathcal{L}_2^{+} & \theta^{-}f_2(H_2^-)
\end{pmatrix},
\label{3.2.1}
\end{equation}
where $f_2$ is a real function. Then, the action of $A_{\mathcal{B}}^{-}$ over the eigenstates (\ref{2.2.14}) is ruled by 
\begin{equation}
A_{\mathcal{B}}^{-}\Psi_n(x,y)=2\sqrt{p_n}f_2(\mathcal{E}_n)\Psi_{n-1}(x,y)\times
\left\lbrace
\begin{array}{c}
0\quad\mbox{for}\quad n=0,\\
\\
\frac{1}{2}\quad\mbox{for}\quad n=1,\\
\\
\frac{1}{\sqrt{2}}\quad\mbox{for}\quad n=2\\
\\
1\quad\mbox{for}\quad n\geq 3.\\
\end{array}
\right.
\label{3.2.2}
\end{equation}
The corresponding creation operator $A_{\mathcal{B}}^{+}\equiv\left(A_{\mathcal{B}}^{-}\right)^{\dagger}$ is
\begin{equation}
A_{\mathcal{B}}^{+}=
\begin{pmatrix} 
\mathcal{L}_2^{-}\frac{f_2(H_2^-)}{\sqrt{(H_2^--\mathcal{E}_0)(H_2^--\mathcal{E}_1)}}\theta^{+}\frac{1}{\sqrt{(H_2^--\mathcal{E}_0)(H_2^--\mathcal{E}_1)}}\mathcal{L}_2^{+} &\mathcal{L}_2^{-}\frac{f_2(H_2^-)}{\sqrt{(H_2^--\mathcal{E}_0)(H_2^--\mathcal{E}_1)}} \theta^{+}\\ 
&\\
f_2(H_2^-)\theta^{+}\frac{1}{\sqrt{(H_2^--\mathcal{E}_0)(H_2^--\mathcal{E}_1)}}\mathcal{L}_2^{+}& f_2(H_2^-)\theta^{+}
\end{pmatrix},
\label{3.2.3}
\end{equation}
which fulfills
\begin{equation}
A_{\mathcal{B}}^{+}\Psi_n(x,y)=2\sqrt{q_n}f_2(\mathcal{E}_{n+1})\Psi_{n+1}(x,y)\times
\left\lbrace
\begin{array}{c}
\frac{1}{2}\quad\mbox{for}\quad n=0,\\
\\
\frac{1}{\sqrt{2}}\quad\mbox{for}\quad n=1,\\
\\
1\quad\mbox{for}\quad n\geq 2.\\
\end{array}
\right.
\label{3.2.4}
\end{equation}
Once again, we are able to construct the $n$-th eigenstate through the successive application of $A_{\mathcal{B}}^+$ over $\Psi_0$. as long as $f_2(\mathcal{E}_{m})\neq 0$ for $m\in[1,n]$.
\section{Generalized coherent states}
In quantum mechanics the coherent states can be constructed through different definitions, which for the harmonic oscillator are mutually equivalent. However, for a system different from the oscillator in general these definitions are not equivalent, thus leading to different sets of coherent states.\\

In the previous section we have found a family of ladder operators for monolayer and bilayer graphene which are well determined, up to an arbitrary function $f_1$ or $f_2$ in each case. In order to work both cases simultaneously, let us define
\begin{equation}
f_1(\mathcal{E}_n)=f(n)\times\left\lbrace
\begin{array}{c}
\frac{1}{\sqrt{2}}\quad\mbox{for}\quad
n=1,\\
\\
\frac{1}{2}\quad\mbox{for}\quad
n\geq2,
\end{array}\right.
\label{4.1}
\end{equation}
and 
\begin{equation}
f_2(\mathcal{E}_n)=f(n)\times\left\lbrace
\begin{array}{c}
1\quad\mbox{for}\quad
n=1,\\
\\
\frac{1}{\sqrt{2}}\quad\mbox{for}\quad
n=2,\\
\\
\frac{1}{2}\quad\mbox{for}\quad
n\geq3,
\end{array}\right.
\label{4.2}
\end{equation}
which translate equations (\ref{3.1.2}) and (\ref{3.2.2}) into
\begin{equation}
A^-_{\mathcal{G}}\Psi_n(x,y;\mathcal{G})=\sqrt{p_n}f(n)\Psi_{n-1}(x,y;\mathcal{G})
\qquad
\mbox{for}
\qquad
n=0,1,\dots
\label{4.3}
\end{equation}
where $\mathcal{G}=\mathcal{M},\mathcal{B}$ and $\Psi_n(x,y;\mathcal{G})$ represents the monolayer or bilayer graphene eigenfunctions. We must emphasize that this relationship is similar only algebraically, since the eigenstates and the annihilation operator in both cases have different nature. Next we will work with the generalized coherent states, based on relationship (\ref{4.3}).
\subsection{Barut–Girardello coherent states}
In the literature the eigenstates of the system annihilation operator are known as
Barut-Girardello coherent states (BGCS)\cite{Barut:1970qf}. Thus, for the annihilation operator of equation (\ref{4.3}) the BGCS must fulfill
\begin{equation}
A^{-}_{\mathcal{G}}\Psi^{\mbox{\tiny{BG}}}_{\alpha}(x,y;\mathcal{G})=\alpha\Psi^{\mbox{\tiny{BG}}}_{\alpha}(x,y;\mathcal{G}),
\label{4.1.1}
\end{equation}
where $\alpha$ is a complex eigenvalue. The BGCS can be represented in the basis of Hamiltonian eigenfunctions as the following linear combination,
\begin{equation}
\Psi^{\mbox{\tiny{BG}}}_{\alpha}(x,y;\mathcal{G})=\sum_{n=0}^{\infty}a_n\Psi_n(x,y;\mathcal{G}).
\label{4.1.2}
\end{equation}
By substituting this expression into equation (\ref{4.1.1}), after some algebra we get that
\begin{equation}
\alpha a_n=f(n+1)\sqrt{p_{n+1}}a_{n+1}\qquad
\mbox{for}\qquad
n=0,1,...
\label{4.1.3}
\end{equation}
The previous recurrence relationship supplies an iterative formula for $a_n$ in which $f$ plays an important role. Moreover, two different cases appear, according to the values that $f(n)$ can take.
\subsubsection{Case with $f(n)\neq 0$}
If $f(n)\neq0$ for $n=1,2,...$ the recurrence relationship (\ref{4.1.3}) leads to
\begin{equation}
a_n=\frac{\alpha^{n}a_0}{\sqrt{\left[p_n\right]!}\left[f(n)\right]!},
\label{4.1.1.1}
\end{equation}
where the generalized factorial function is defined by
\begin{equation}
\left[f(k)\right]!\equiv\left\lbrace
\begin{array}{c}
1\quad\mbox{for}\quad k=0,\\
\\
f(1)f(2)\cdots f(k)
\quad\mbox{for}\quad k=1,2,...
\end{array}
\right.
\label{4.1.1.2}
\end{equation}
Then, the normalized BGCS for this case turn out to be 
\begin{equation}
\Psi^{\mbox{\tiny{BG}}}_{\alpha}(x,y;\mathcal{G})=\left[\sum_{n=0}^{\infty}\frac{|\alpha|^{2n}}{\left[p_n\right]!\left(\left[f(n)\right]!\right)^2}\right]^{-\frac{1}{2}}\left[\sum_{n=0}^{\infty}\frac{\alpha^{n}}{\sqrt{\left[p_n\right]!}\left[f(n)\right]!}\Psi_n(x,y;\mathcal{G})\right].
\label{4.1.1.3}
\end{equation}
Despite $f$ is an arbitrary function, we must restrict ourselves to functions allowing the convergence of our coherent state expression (\ref{4.1.1.3}).
\subsubsection{Case with $f(n)=0$ for some $n$}
In this case we have to take into account that $f$ has roots for some $n\in\mathbb{N}$; let us denote as $m$ the maximum value of $n$ for which there is a root, with $m\geq 1$. Note that it does not matter for the analysis if either there are more roots or not, the conditions we get turn out to be
\begin{equation}
a_n=0\quad\mbox{for}\quad
0 \leq n \leq m-1,
\label{4.1.2.1}
\end{equation}
\begin{equation}
a_{n+1}=\frac{\alpha a_n}{\sqrt{p_{n+1}}f(n+1)}
\quad\mbox{for}\quad
n\geq m.
\label{4.1.2.2}
\end{equation}
Thus, for this case the normalized BGCS are
\begin{equation}
\Psi^{\mbox{\tiny{BG}}}_{\alpha}(x,y;\mathcal{G})=\left[\sum_{n=0}^{\infty}\frac{|\alpha|^{2n}}{\left[\hat{p}_n\right]!\left(\left[\hat{f}(n)\right]!\right)^2}\right]^{-\frac{1}{2}}\left[\sum_{n=0}^{\infty}\frac{\alpha^{n}}{\sqrt{\left[\hat{p}_n\right]!}\left[\hat{f}(n)\right]!}\Psi_{n+m}(x,y;\mathcal{G})\right].
\label{4.1.2.3}
\end{equation}
Two points are worth to be stressed: firstly, the linear combination starts from $\Psi_m(x,y;\mathcal{G})$; secondly, the coefficients in such decomposition have changed as compared with equation (\ref{4.1.1.3}), since now we have $\hat{p}_n\equiv p_{n+m}$ and $\hat{f}(n)\equiv f(n+m)$.
Note that in both cases the coherent states look similar to the ones derived in \cite{Diaz-Bautista2017,Fernandez2020}. In fact a proper choice of $f$ allows us to recover them. Nevertheless, the way we have constructed the annihilation operator and the form we have fixed the function $f$ will lead us to new interesting properties.
\subsection{Gilmore-Perelomov coherent states}
The displacement operator for the harmonic oscillator allows to construct the coherent states straightforwardly,
but an important fact in that construction is the really simple commutation relationship between the annihilation and creation operators. For an arbitrary system with defined ladder operators it arises the idea that the coherent states can be obtained as the result of acting a generalized displacement operator on one extremal state of the system (a state annihilated by the operator $A_{\mathcal{G}}^{-}$). In the literature these are called Gilmore-Perelomov coherent states (GPCS) \cite{cmp/1103858078,gilmore2012lie}. Next, we carry out first the analysis of the commutator $\left[A_{\mathcal{G}}^{-},A_{\mathcal{G}}^{+}\right]$ and then generate the Gilmore-Perelomov coherent states.\\

Let us begin assuming that equations (\ref{4.1}) and (\ref{4.2}) are fulfilled, thus we have
\begin{align}
A_{\mathcal{G}}^{-}\Psi_n(x,y;\mathcal{G})&=\sqrt{p_n}f(n)\Psi_{n-1}(x,y;\mathcal{G}),
\label{4.2.1}\\
A_{\mathcal{G}}^{+}\Psi_n(x,y;\mathcal{G})&=\sqrt{q_n}f(n+1)\Psi_{n+1}(x,y;\mathcal{G})\quad\forall
\;n.
\label{4.2.2}
\end{align}
We are interested in knowing if we can get the commutation relationship $\left[A_{\mathcal{G}}^{-},A_{\mathcal{G}}^{+}\right]=1$, as for the harmonic oscillator. For doing this, we use an arbitrary state expressed as a linear combination of the basis vectors, namely $\Psi(x,y;\mathcal{G})=\sum_{n=0}^{\infty}a_n\Psi_n(x,y;\mathcal{G})$. Thus,
\begin{equation}
\left[A_{\mathcal{G}}^{-},A_{\mathcal{G}}^{+}\right]\Psi(x,y;\mathcal{G})=\sum_{n=0}^{\infty}a_n\left(\gamma_{n+1}-\gamma_{n}\right)\Psi_n(x,y;\mathcal{G}),
\label{4.2.3}
\end{equation}
where $\gamma_n$ is a real non-negative function of $n$ given by 
\begin{equation}
\gamma_n=\sqrt{q_{n-1}p_n}f^2(n).
\label{4.2.4}
\end{equation}
Once again we get two different cases, according to the values taken by $f$.
\subsubsection{Case with $f(n)\neq 0$}
If $f$ has no roots for $n\in\mathbb{N}$, then the only extremal state will be the ground state $\Psi_0(x,y;\mathcal{G})$. The ladder operators commute to the identity if the following conditions are fulfilled
\begin{align}
f(1)=&\left(q_0p_1\right)^{-\frac{1}{4}},\label{4.2.1.1}\\
f(n+1)=&\left(\frac{1+\sqrt{q_{n-1} p_{n}}f^{2}(n)}{\sqrt{q_n p_{n+1}}}\right)^{\frac{1}{2}}
\quad\mbox{for}\quad
n\geq 1.
\label{4.2.1.2}
\end{align}
Equations (\ref{4.2.1.1}) and (\ref{4.2.1.2}) ensure that $\left[A^-_{\mathcal{G}},A^+_{\mathcal{G}}\right]=1$, i.e., the operators $A^-_{\mathcal{G}}$ and $A^+_{\mathcal{G}}$ fulfill the Heisenberg-Weyl algebra. Then, the displacement operator defined by $D_{\mathcal{G}}\left(\alpha\right)=\exp \left(\alpha A_{\mathcal{G}}^{+}-\alpha ^{\ast }A_{\mathcal{G}}^{-}\right)$ can be factorized using the Baker-Hausdorff formula as $D_{\mathcal{G}}(\alpha )=e^{-\frac{1}{2}|\alpha|^{2}}e^{\alpha A_{\mathcal{G}}^{+}}e^{-\alpha^{*}A_{\mathcal{G}}^{-}}$.

The Gilmore-Perelomov coherent states $\Psi^{\mbox{\tiny{GP}}}_{\alpha}(x,y;\mathcal{G})$ in this case are the result of acting $D_{\mathcal{G}}(\alpha)$ over the ground state (the only extremal state in this case), thus we have
\begin{equation}
\Psi^{\mbox{\tiny{GP}}}_{\alpha}(x,y;\mathcal{G})=e^{-\frac{1}{2}|\alpha|^{2}}\sum_{n=0}^{\infty}
\frac{\alpha^{n}\sqrt{\left[\hat{q}_n\right]!}\left[f(n)\right]!}{n!}
\Psi_n(x,y;\mathcal{G}),
\label{4.2.1.3}
\end{equation}
where $\hat{q}_n=q_{n-1}$. It is important to remark that here we do not need to normalize the Gilmore-Perelomov coherent states, since $D_{\mathcal{G}}(\alpha)$ is a unitary operator.
\subsubsection{Case with $f(n)=0$ for some $n$}
Let us start by considering the ordered set of integers $\left\lbrace m_i|f(m_i)=0\quad\mbox{for} \quad i=1,2,...,l\right\rbrace$, i.e., the extremal states are $\Psi_{m_i}(x,y;\mathcal{G})$. Notice that $\gamma_{m_i}=0$ since $f(m_i)=0$, which implies that the coefficients multiplying $\Psi_{m_i}(x,y;\mathcal{G})$ and $\Psi_{m_i-1}(x,y;\mathcal{G})$ in equation (\ref{4.2.3}) reduce to $a_{m_i}\gamma_{m_i+1}$ and $-a_{m_i-1}\gamma_{m_i-1}$ respectively. While the first coefficient can be made equal to $a_{m_i}$ for some particular election of $f$, the second one can not be made equal to $a_{m_i-1}$ because $-\gamma_{m_i-1}$ is non-positive. As a consequence, $\left[A^-_{\mathcal{G}},A^+_{\mathcal{G}}\right]$ can not be equal to the identity operator in this case, under any circumstances.

Although now $\left[A^-_{\mathcal{G}},A^+_{\mathcal{G}}\right]\neq 1$, there are still some cases that may be worth of some discussion.

a) If $m_{i+1}-m_{i}\geq 2$ for $i=1,2,...$, which means that all the roots of $f$ are non-consecutive, thus under an appropriate choice of $f$ equation (\ref{4.2.3}) might produce a slight modification on the contribution of the extremal states $\Psi_{m_i}(x,y;\mathcal{G})$ to the linear combination, i.e., the commutator could be made very close to the identity.

b) If $m_{i+1}-m_{i}=1$ for some $i$, at least two roots of $f$ are consecutive which implies that the term $\Psi_{m_i}(x,y;\mathcal{G})$ will disappear from the linear combination (\ref{4.2.3}). Then, the commutator $\left[A_{\mathcal{G}}^{-},A_{\mathcal{G}}^{+}\right]$ is very close to the identity, minus the projector onto the subspace generated by $\Psi_{m_i}(x,y;\mathcal{G})$.

The most important consequence of having roots in $f(n)$ is that the Baker-Hausdorff factorization formula can not be applied anymore. Nevertheless, we can use an alternative non-unitary displacement operator defined by $D_{\mathcal{G}}(\alpha)=e^{\alpha A_{\mathcal{G}}^{+}}$. The Gilmore-Perelomov coherent states arise from acting this operator over the extremal states $\Psi_{m_i}(x,y;\mathcal{G})$. By remembering that there are $l$ roots for $f$, the coherent states are thus given by
\begin{equation}
\Psi^{\mbox{\tiny{GP}}}_{\alpha}(x,y,m_i;\mathcal{G})=\sum_{n=0}^{m_{i+1}-m_i-1}\frac{\alpha^n\sqrt{\left[\tilde{q}(n)\right]!}[\tilde{f}(n)]!}{n!}
\Psi_{n+m_i}(x,y;\mathcal{G}),\quad
\mbox{for}\quad
i=0,1,2,...,l,
\label{4.2.2.1}
\end{equation}
where $m_0\equiv0$, $m_{l+1}\equiv\infty$, $\tilde{q}(n)=q(n+m_i-1)$ and $\tilde{f}(n)=f(n+m_i)$. Some important points need to be stressed: the GPCS in this case have two labels, the complex number $\alpha$ and the extremal state label $m_i$ they are generated from; these states are orthogonal with respect to the last label. In addition, since the displacement operator employed is non-unitary, the coherent states of equation (\ref{4.2.2.1}) should still be normalized.
\subsection{Minimum uncertainty coherent states}
Another definition has been promoted by Aragone, Nieto and Simmons \cite{Schrodinger1926,PhysRevLett.41.207,PhysRevD.20.1321}, in which the coherent state $\Psi^{\mbox{\tiny{MU}}}(x,y;\mathcal{G})$, called minimum uncertainty coherent state, must saturate the uncertainty product for the quadratures $Q_{\mathcal{G}}$ and $P_{\mathcal{G}}$, defined by
\begin{equation}
Q_{\mathcal{G}}=\frac{1}{\sqrt{2}}\left(A_{\mathcal{G}}^{+}+A_{\mathcal{G}}^{-}\right),
\qquad
P_{\mathcal{G}}=\frac{i}{\sqrt{2}}\left(A_{\mathcal{G}}^{+}-A_{\mathcal{G}}^{-}\right).
\label{4.3.1}
\end{equation}
Note that, if two arbitrary operators $F$ and $G$ are hermitian, their commutator $\left[F,G\right]$ should be antihermitian, so it can be defined $\left[F,G\right]=iK $, where $K$ has to be hermitian. Thus, for an arbitrary state $\ket{\Psi}$ it must hold that
\begin{equation}
\Delta F\Delta G\geq\frac{1}{2}|\braket{K}|,
\label{4.3.2}
\end{equation} 
where $\braket{K}=\bra{\Psi}K\ket{\Psi}$ and $(\Delta K)^{2}\equiv\bra{\Psi}K^2\ket{\Psi}-\bra{\Psi}K\ket{\Psi}^2$. Equation (\ref{4.3.2}) will be saturated if it is fulfilled
\begin{equation}
\left(F+i\frac{\Delta F}{\Delta G} G\right)\Psi=\left(\braket{F}+i\frac{\Delta F}{\Delta G} \braket{G}\right)\Psi.
\label{4.3.3}
\end{equation}  
Finally, if we make $F=Q_\mathcal{G}$ and $G=P_\mathcal{G}$, after some algebra the previous condition becomes
\begin{equation}
\left[\frac{(1-\lambda)}{(1+\lambda)}A_\mathcal{G}^{+}+A_\mathcal{G}^{-}\right]\Psi=\alpha\Psi,
\label{4.3.4}
\end{equation}
where $\lambda=\Delta Q_\mathcal{G}/\Delta P_\mathcal{G}$ and $\alpha=\left[\frac{(1-\lambda)}{(1+\lambda)}\braket{A_\mathcal{G}^{+}}+\braket{A_\mathcal{G}^{-}}\right]$.

Some points are worth to be stressed: for this treatment the quadratures $Q_{\mathcal{G}}$ and $P_{\mathcal{G}}$ are not, in general, the position and moment operators. On the other hand, the previous equation represents the eigenvalue problem for the so-called squeezed coherent states, which saturate the Heisenberg uncertainty product $\Delta Q_\mathcal{G} \Delta P_\mathcal{G}$ but the uncertainty in one of their quadratures is in general different from the other one. The particular case with $\lambda=1$ ($\Delta Q_\mathcal{G}=\Delta P_\mathcal{G}$) reduces to an equation of eigenvalues for $A_{\mathcal{G}}^{-}$, therefore the minimum uncertainty coherent states turn out to be equal to the  Barut-Girardello coherent states, i.e., $\Psi^{\mbox{\tiny{MU}}}(x,y;\mathcal{G})=\Psi^{\mbox{\tiny{BG}}}(x,y;\mathcal{G})$; from now on we will refer to such a case as the minimum uncertainty coherent states.
\section{Constant magnetic field}
Once we have discussed in general the coherent states for monolayer and bilayer graphene, we are going to analyze now the particular case of a constant magnetic field, for which we will compare the results of the different definitions and some of their associated physical quantities.

Let us consider the constant magnetic field $\vec{B}=\mathfrak{B}_{0}\hat{e}_z$ with $\mathfrak{B}_0>0$, so the vector potential amplitude is $\mathcal{A}(x)=\mathfrak{B}_0x$. Then, the eigenfunctions $\psi^{-}_n(x)$ of $H^{-}_i$, $i=1,2$ for both monolayer and bilayer graphene are given as follows:
\begin{equation}
\psi^{-}_n(x)=\sqrt{\frac{1}{2^n n!}\left(\frac{\omega}{2\pi}\right)^{\frac{1}{2}}}H_n\left[\sqrt{\frac{\omega}{2}}\left(x+\frac{2k}{\omega}\right)\right]e^{-\frac{\omega}{4}\left(x+\frac{2k}{\omega}\right)^{2}},\qquad
n=0,1...
\label{5.1}
\end{equation}
where $\omega\equiv2e\mathfrak{B}_0/c\hbar$ and $k$ is the wave number in $y$ direction. The corresponding eigenvalues are $\mathcal{E}_n=n\omega$ (see \cite{Kuru_2009,Fern_ndez_C_2020,Midya_2014,Fernandez_2021}). Thus, the eigenfunctions and eigenvalues for monolayer graphene become
\begin{equation}
\Psi_n(x,y;\mathcal{M})=\frac{1}{\sqrt{2^{1-\delta_{n0}}}}e^{iky}
\begin{pmatrix} 
(1-\delta_{n0})\psi_{n-1}^-(x)\\ 
i\psi_n^-(x)
\end{pmatrix},
\quad
E_n=\hbar v_{{\scriptscriptstyle F}}\sqrt{n\omega},\qquad n=0,1,...
\label{5.2}
\end{equation}
while for bilayer graphene are
\begin{equation}
\Psi_n(x,y;\mathcal{B})=\frac{e^{iky}}{\sqrt{2^{(1-\delta_{n0}-\delta_{n1})}}}
\begin{pmatrix} 
(1-\delta_{n0}-\delta_{n1})\psi_{n-2}^-(x)\\ 
\psi_n^-(x)
\end{pmatrix},
\qquad
E_n=\frac{\hbar^2 \omega}{2m^*}\sqrt{n(n-1)},\quad n=0,1,...
\label{5.3}
\end{equation}
Since the involved eigenfunctions correspond to the harmonic oscillator, it is required the one-dimensional first-order ladder operators given by 
\begin{align}
\theta^{-}&=\frac{1}{\sqrt{2}}\left(z+\frac{d}{dz}\right),\nonumber\\
\theta^{+}&=\frac{1}{\sqrt{2}}\left(z-\frac{d}{dz}\right),
\label{5.4}
\end{align}
where $z=\sqrt{\frac{\omega}{2}}\left(x+\frac{2k}{\omega}\right)$, such that
\begin{align}
\theta^{-}\psi_n^{-}&=\sqrt{n}\psi_{n-1}^{-},\nonumber\\
\theta^{+}\psi_n^{-}&=\sqrt{n+1}\psi_{n+1}^{-},\qquad\mbox{for}\qquad n=0,1,...
\label{5.5}
\end{align}
Equations (\ref{5.1}-\ref{5.5}) are all we need to determine the generalized coherent states. From now on we will restrict ourselves to the case where the function $f$ has no roots, in particular, we will choose $f(n)=1$ for the BGCS of equation (\ref{4.1.1.3}).
\subsection{Generalized coherent states}
Once we have fixed $f(n)=1$, looking at equations (\ref{3.1}), (\ref{5.4}) we can simplify the BGCS of equation (\ref{4.1.1.3}),
which in our approach coincide with the MUCS, i.e.,
\begin{align}
\Psi^{\mbox{\tiny{BG}}}_{\alpha}(x,y;\mathcal{G})&=\Psi^{\mbox{\tiny{MU}}}_{\alpha}(x,y;\mathcal{G})\nonumber\\
&=e^{-\frac{1}{2}|\alpha|^{2}}\sum_{n=0}^{\infty}
\frac{\alpha^{n}}{\sqrt{n}!}
\Psi_n(x,y;\mathcal{G}),
\label{5.1.1}
\end{align}
where the index $\mathcal{G}=\mathcal{M},\mathcal{B}$ indicates monolayer or bilayer graphene respectively. Concerning the GPCS, we have to simplify equation (\ref{4.2.1.2}) by using equation (\ref{5.5}). After some algebra we get
\begin{align}
\Psi^{\mbox{\tiny{GP}}}_{\alpha}(x,y;\mathcal{G})&=e^{-\frac{1}{2}|\alpha|^{2}}\sum_{n=0}^{\infty}
\frac{\alpha^{n}}{\sqrt{n}!}
\Psi_n(x,y;\mathcal{G})\nonumber\\
&=\Psi^{\mbox{\tiny{BG}}}_{\alpha}(x,y;\mathcal{G})=\Psi^{\mbox{\tiny{MU}}}_{\alpha}(x,y;\mathcal{G}).
\label{5.1.2}
\end{align}
Equations (\ref{5.1.1}) and (\ref{5.1.2}) indicate the existence of a family of coherent states for which the three definitions are mutually equivalent, which is achieved through the function $f$ such that $f(n)=1$, $n=1,2,...$ It is important to note that $f(n)=1$ does not mean necessarily that $f(x)=1, \forall x\in\mathbb{R}$, it is enough that the function takes just the unit value for any positive integer. Therefore, there are different creation and annihilation operators in equations (\ref{3.1.1}), (\ref{3.1.3}) and (\ref{3.2.1}), (\ref{3.2.3}) leading to the same set of coherent states.

Due to the mutual equivalence between these definitions of generalized coherent states, from now on we will just call them coherent states, and denote them as $\Psi_{\alpha}(x,y;\mathcal{G})$. We must emphasize that these states form an overcomplete set of vectors in the Hilbert space which fulfill a resolution of the identity as follows
\begin{equation}
\mathbb{I}=\frac{1}{\pi}\int_{\mathbb{C}}\ket{\Psi_{\alpha}}\bra{\Psi_{\alpha}}d\mu(\alpha),
\label{5.1.3}
\end{equation}
where $d\mu(\alpha)=\frac{rdrd\theta}{\pi}$ and $\alpha=re^{i\theta}$. In the following sections we will calculate some physical quantities associated with these coherent states.
\subsubsection{Probability density}
An important quantity to be analyzed for these coherent states is the probability density, which is defined as
\begin{equation}
\rho_{\mathcal{G}}(x,y;\alpha)=\Psi_{\alpha}^{\dagger}(x,y;\mathcal{G})\Psi_{\alpha}(x,y;\mathcal{G}).
\label{5.2.1}
\end{equation}
In order to simplify the expression resulting from equation (\ref{5.1.1}) let us define
\begin{equation}
a_n(\alpha)\equiv e^{-\frac{1}{2}|\alpha|^{2}}\frac{\alpha^{n}}{\sqrt{n}!},\qquad
\rho_{n,m}(x,y;\mathcal{G})\equiv\Psi_m(x,y;\mathcal{G})^{\dagger}\Psi_n(x,y;\mathcal{G}),
\nonumber
\end{equation}
\begin{equation}
C_{m,n}(x,y,\alpha;\mathcal{G})=a_m^{*}(\alpha)a_n(\alpha)\rho_{n,m}(x,y;\mathcal{G}),
\label{5.2.2}
\end{equation}
with $\rho_{n,m}$ being symmetric under the exchange of indexes, due to the real valued functions involved in equation (\ref{5.1}). Notice that $\rho_{n,m}(x,y;\mathcal{G})=\rho_{n,m}(x;\mathcal{G})$, i.e., the probability density depends only on $x$ in the way 
\begin{equation}
\rho_{\mathcal{G}}(x;\alpha)=\sum^{\infty}_{n,m=0}\mathfrak{Re}\left(C_{m,n}(x,\alpha;\mathcal{G})\right).
\label{5.2.3}
\end{equation}
By defining 
\begin{equation}
\rho_n(x;\mathcal{G})\equiv\rho_{n,n}(x;\mathcal{G}),
\label{5.2.4}
\end{equation}
and using the eigenstates (\ref{5.2}) and (\ref{5.3}) respectively, we finally get that
\begin{equation}
\rho_{\mathcal{M}}(x,\alpha)=e^{-r^2}\left\lbrace
\rho_0(x;\mathcal{M})+
2\sum_{n=1}^{\infty}\frac{r^n\mbox{cos}(n\theta)}{\sqrt{n!}}\rho_{n,0}(x;\mathcal{M})+
\sum_{n,m=1}^{\infty}\frac{r^{n+m}\mbox{cos}\left[(n-m)\theta\right]}{\sqrt{n!m!}}\rho_{n,m}(x;\mathcal{M})
\right\rbrace,
\label{5.2.5}
\end{equation}
for monolayer graphene coherent states and 
\begin{align}
\rho_{\mathcal{B}}(x,\alpha)=&e^{-r^2}\left\lbrace
\sum_{n,m=2}^{\infty}\frac{r^{n+m}\mbox{cos}\left[(n-m)\theta\right]}{\sqrt{n!m!}}\rho_{n,m}(x;\mathcal{B})+
\rho_0(x;\mathcal{B})+
r^{2}\rho_1(x;\mathcal{B})+
2r\mbox{cos}(\theta)\rho_{1,0}(x;\mathcal{B})\right.\nonumber\\
&+\left.2\sum_{n=2}^{\infty}\frac{r^n}{\sqrt{n!}}\left[\mbox{cos}(n\theta)\rho_{n,0}(x;\mathcal{B})+r\mbox{cos}\left[(n-1)\theta\right]
\rho_{n,1}(x;\mathcal{B})\right]
\right\rbrace,
\label{5.2.6}
\end{align}
for bilayer graphene coherent states. 

\begin{figure}[ht]
\begin{center}
\subfigure[]{\includegraphics[width=8cm, height=5.7cm]{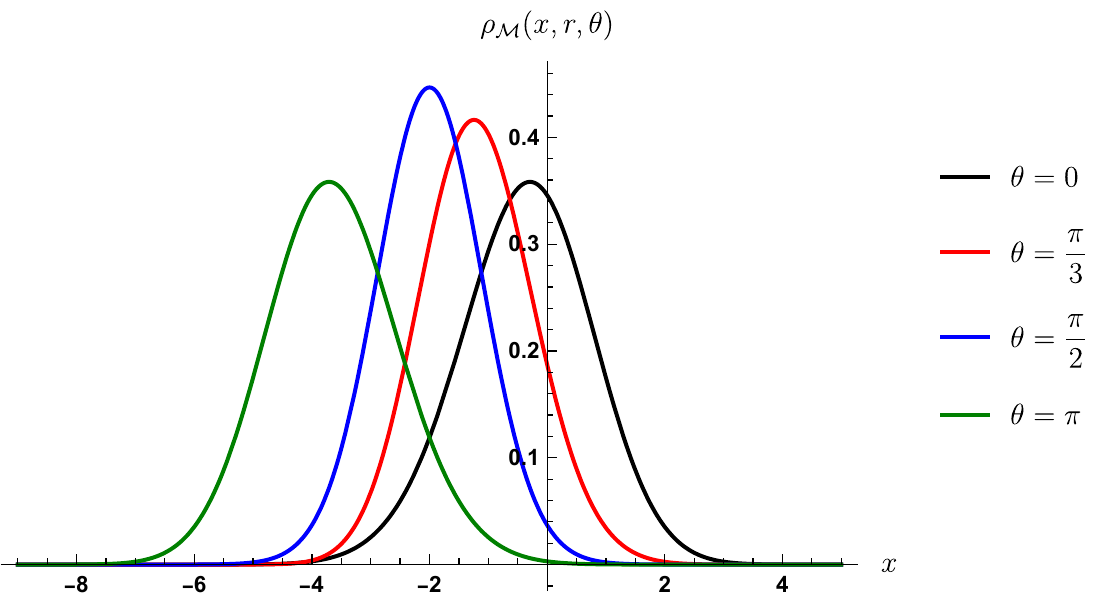}}
\subfigure[]{\includegraphics[width=8cm, height=5.7cm]{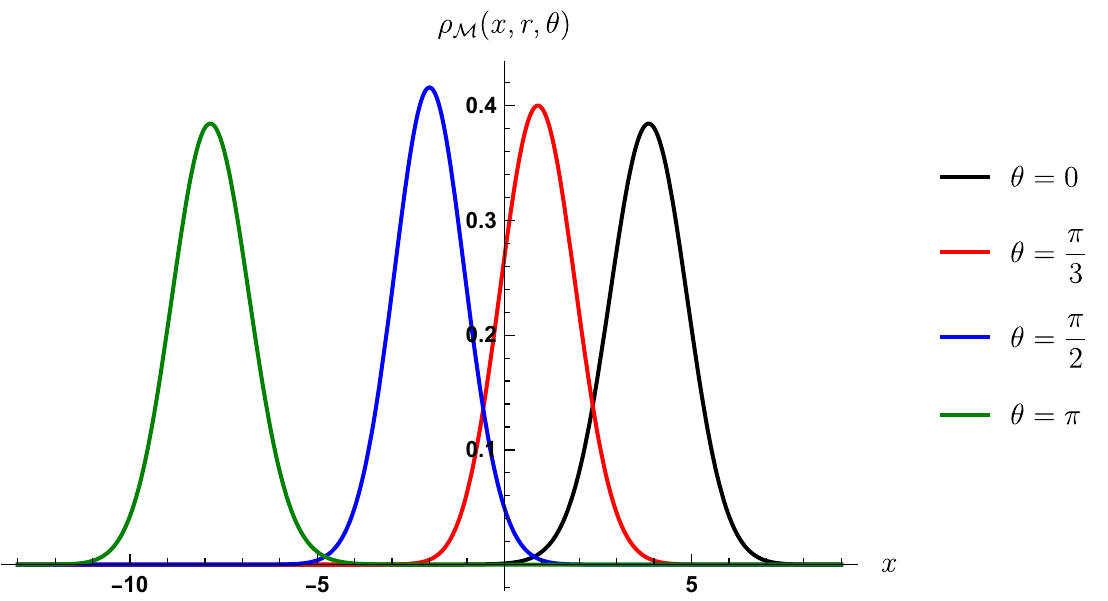}}
\subfigure[]{\includegraphics[width=8cm, height=5.7cm]{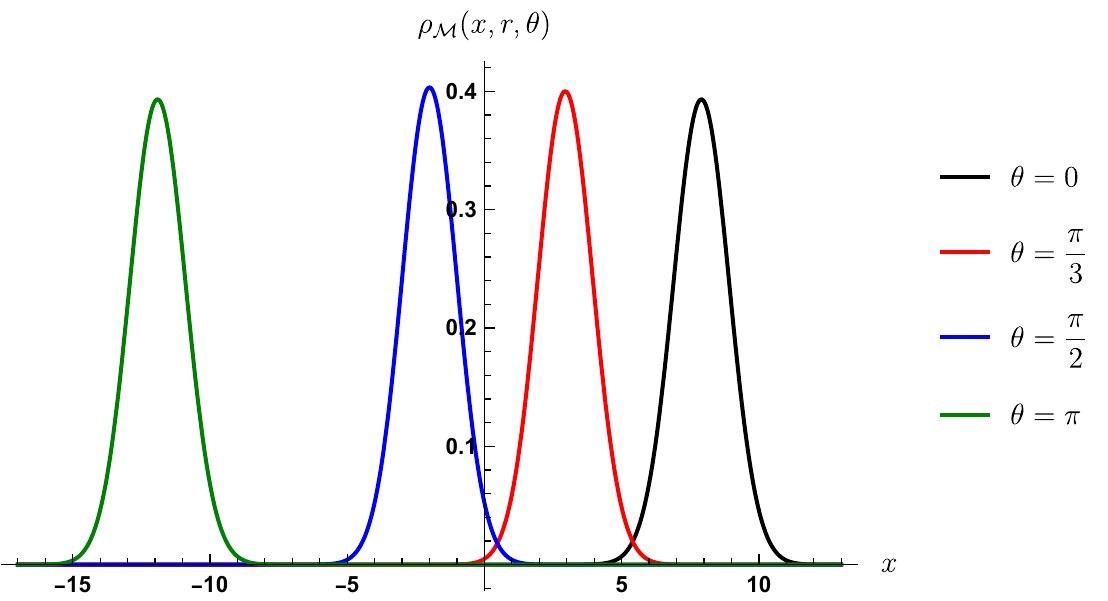}}
\caption{Plots of the probability density for the monolayer graphene coherent states (\ref{5.1.1}) as functions of $x$ and $\theta$, for fixed values of $\omega=k=1$ and different values of $r$: (a) $r=1$; (b) $r=3$; (c) $r=5$. We can see that the maximum shifts to the left as we increase $\theta$.}
\label{F1}
\end{center}
\end{figure}

\begin{figure}[ht] 
\begin{center}
\subfigure[]{\includegraphics[width=8cm, height=5.7cm]{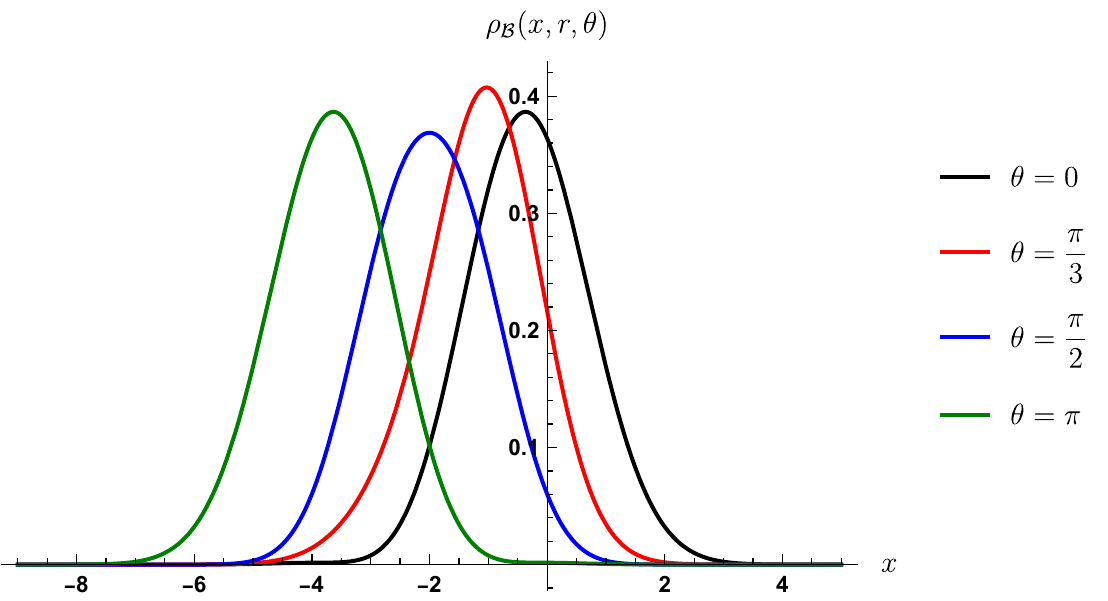}}
\subfigure[]{\includegraphics[width=8cm, height=5.7cm]{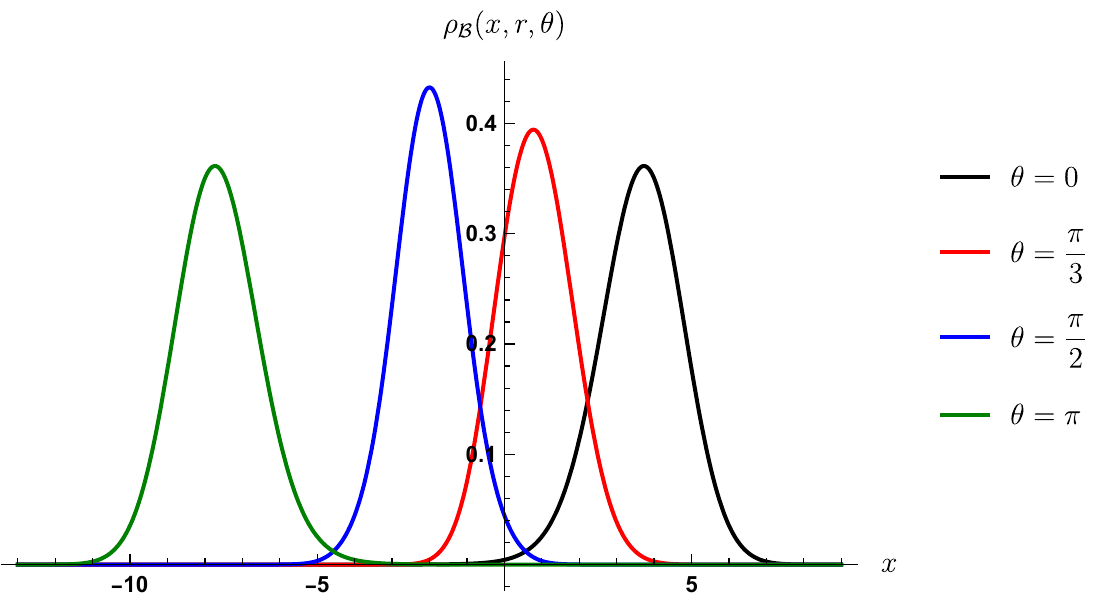}}
\subfigure[]{\includegraphics[width=8cm, height=5.7cm]{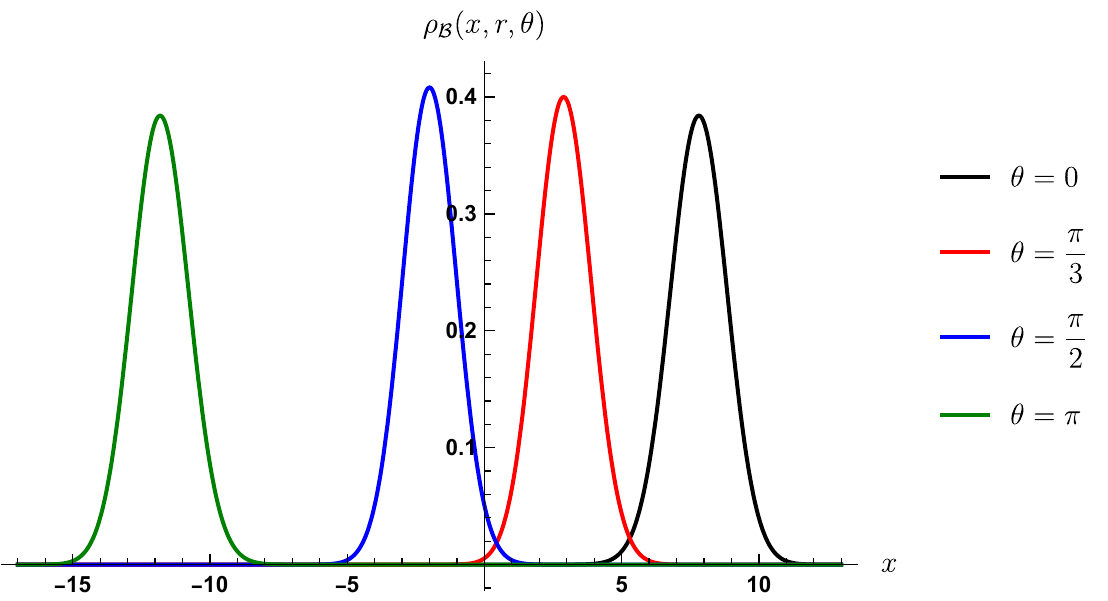}}
\caption{Plots of the probability density for the bilayer graphene coherent states (\ref{5.1.1}) as functions of $x$ and $\theta$, for fixed values of $\omega=k=1$ and different values of $r$: (a) $r=1$; (b) $r=3$; (c) $r=5$. We can see that the maximum shifts to the left as we increase $\theta$.}
\label{F2}
\end{center}
\end{figure}

Note that the change $\theta\rightarrow-\theta$ keeps invariant the probability density, which can be represented as $\rho_{\mathcal{G}}(x,\alpha)=\rho_{\mathcal{G}}(x,\bar{\alpha})$. For a fixed value of $r$, it can be observed that the amplitude and the position of the probability density maximum in both cases depend on the value of $\theta$, as it is shown in Figure \ref{F1} for monolayer graphene and in Figure \ref{F2} for bilayer graphene. 
\newpage
\subsubsection{Current density}
The current expressions for monolayer and bilayer graphene, equations (\ref{11}) and (\ref{2.2.15}) respectively, are different; thus we must treat them separately.

\begin{figure}[ht] 
\begin{center}
\subfigure[]{\includegraphics[width=8cm, height=5.7cm]{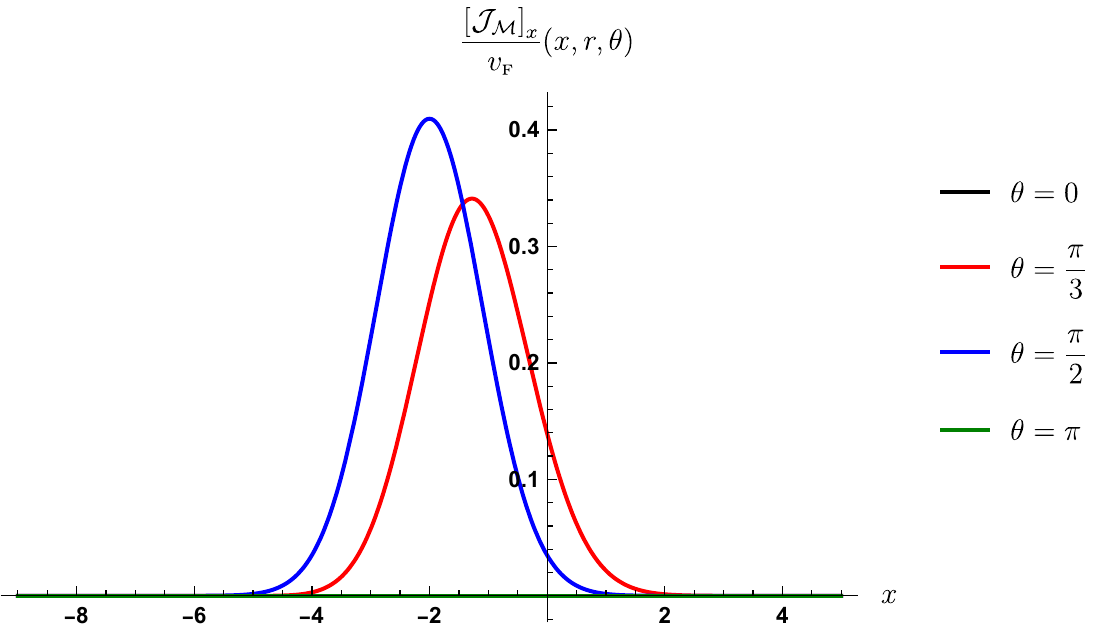}}
\subfigure[]{\includegraphics[width=8cm, height=5.7cm]{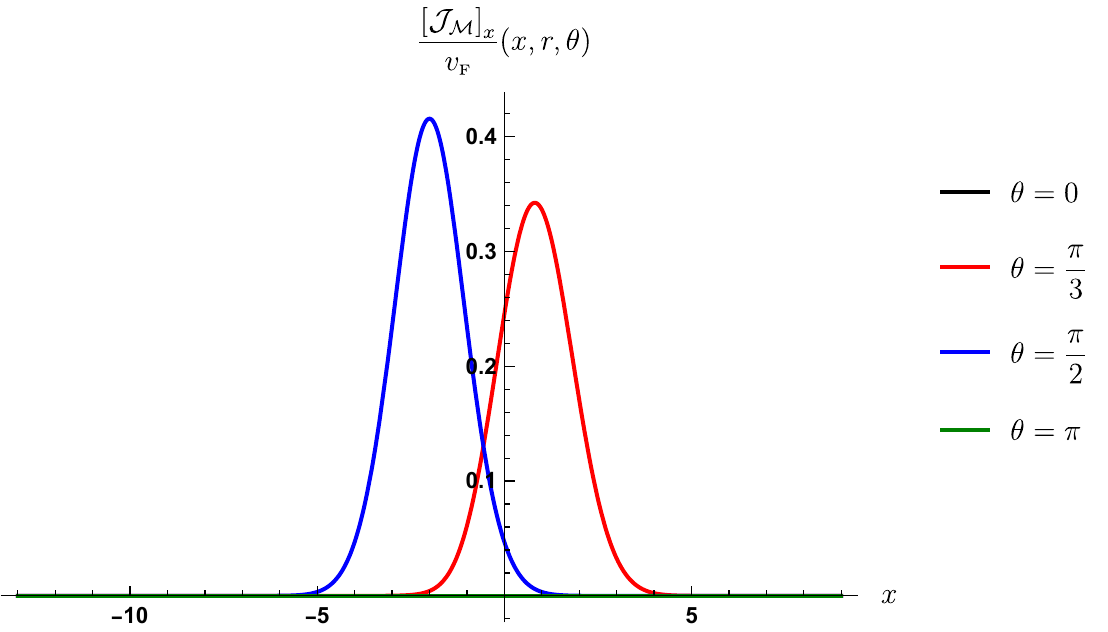}}
\subfigure[]{\includegraphics[width=8cm, height=5.7cm]{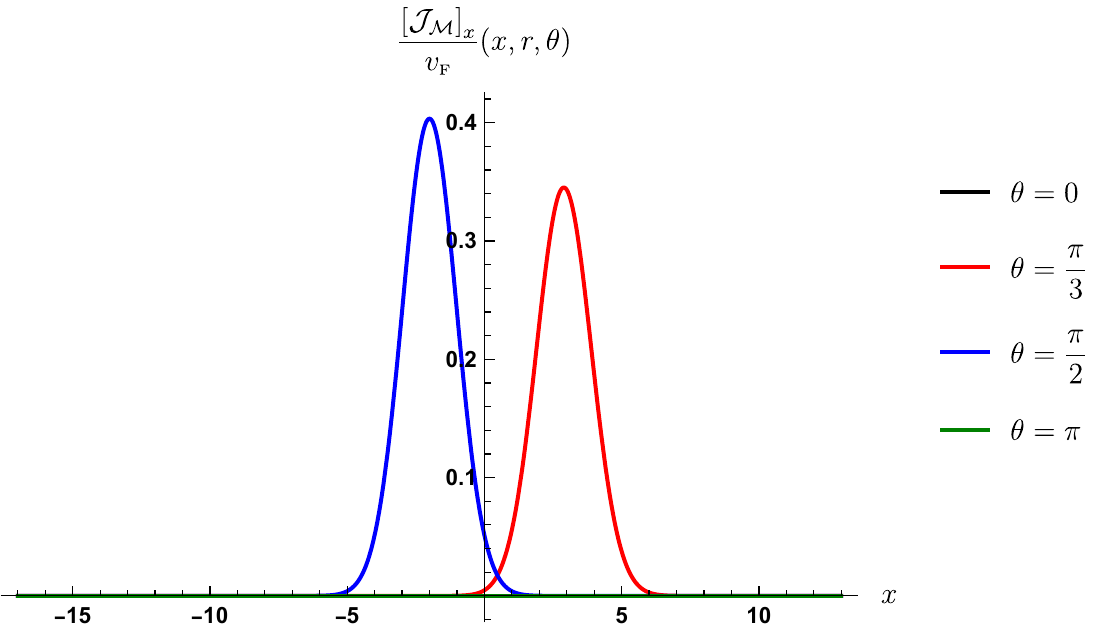}}
\caption{Plots of the $x-$component current density $\left[\mathcal{J}_\mathcal{M}\right]_x$ for the monolayer graphene coherent states (\ref{5.1.1}) as function of $x$ and $\theta$ for fixed values of $\omega=k=1$ and different values of $r$: (a) $r=1$; (b) $r=3$; (c) $r=5$.}
\label{F3}
\end{center}
\end{figure}

\begin{figure}[ht] 
\begin{center}
\subfigure[]{\includegraphics[width=8cm, height=5.7cm]{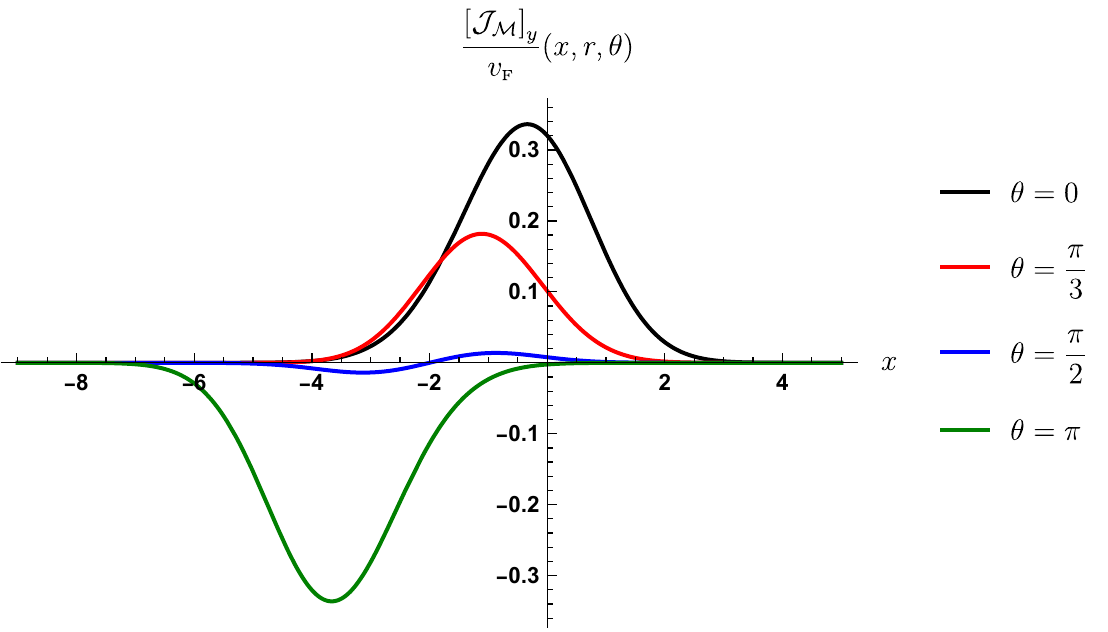}}
\subfigure[]{\includegraphics[width=8cm, height=5.7cm]{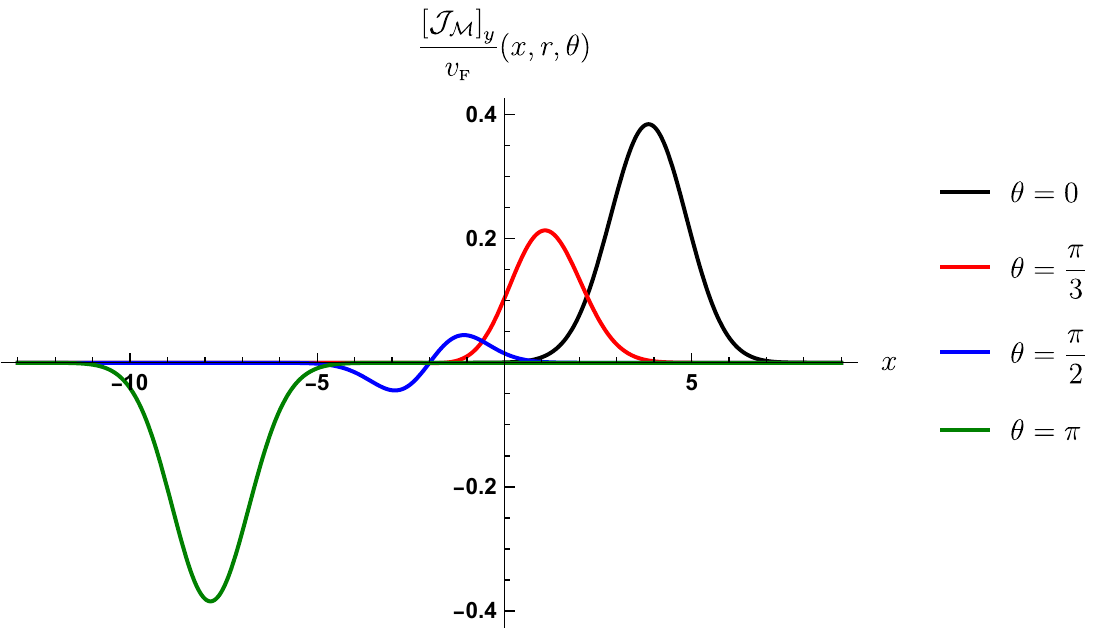}}
\subfigure[]{\includegraphics[width=8cm, height=5.7cm]{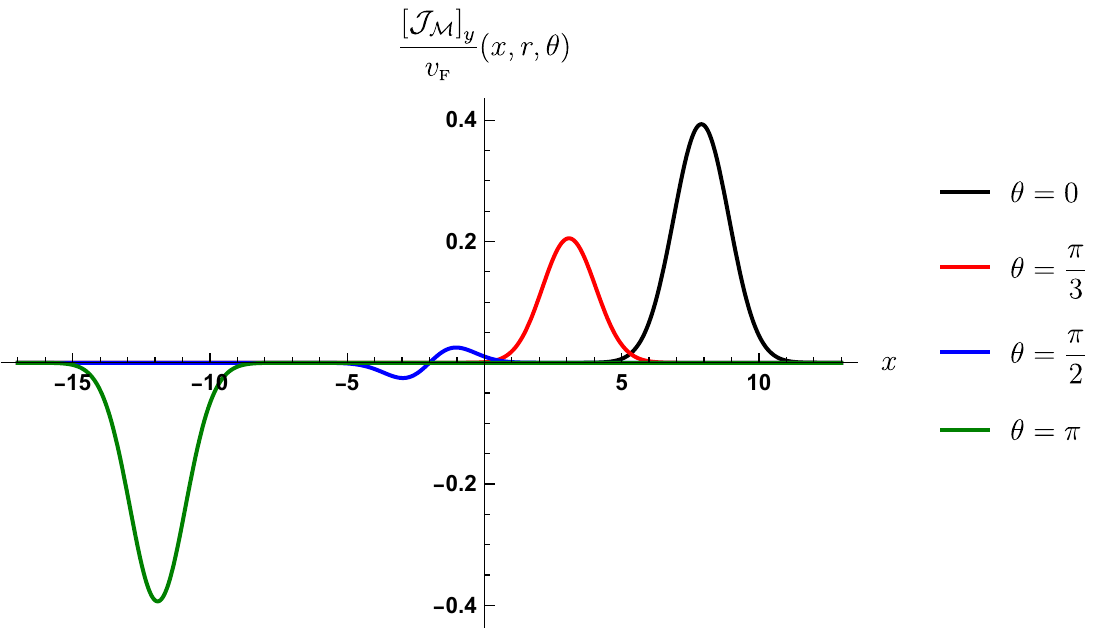}}
\caption{Plots of the $y-$component current density $\left[\mathcal{J}_\mathcal{M}\right]_y$ for the monolayer graphene coherent states (\ref{5.1.1}) as function of $x$ and $\theta$ for fixed values of $\omega=k=1$ and different values of $r$: (a) $r=1$; (b) $r=3$; (c) $r=5$.}
\label{F4}
\end{center}
\end{figure}

For monolayer graphene the current components for the coherent state (\ref{5.1.1}) are given by 
\begin{align}
\left[\vec{\mathcal{J}}_{\mathcal{M}}\right]_x&=v_{{\scriptscriptstyle F}}
e^{-r^2}
\sum_{n,m=0}^{\infty}\frac{r^{n+m}\mbox{sin}\left[(n-m)\theta\right]}{\sqrt{n!m!}}j^{-}_{n,m}(x),\nonumber\\
\left[\vec{\mathcal{J}}_{\mathcal{M}}\right]_y&=v_{{\scriptscriptstyle F}}
e^{-r^2}
\sum_{n,m=0}^{\infty}\frac{r^{n+m}\mbox{cos}\left[(n-m)\theta\right]}{\sqrt{n!m!}}j^{+}_{n,m}(x),
\label{5.3.1}
\end{align}
where 
\begin{equation}
j^{\pm}_{n,m}(x)=\frac{\left(1-\delta_{n,0}\right)\psi^{-}_{n-1}(x)\psi^{-}_{m}(x)
\pm\left(1-\delta_{m,0}\right)\psi^{-}_{n}(x)\psi^{-}_{m-1}(x)}{\sqrt{2^{2-\delta_{n,0}-\delta_{m,0}}}}.
\label{5.3.2}
\end{equation}
By making now $\theta\rightarrow -\theta$ ($\alpha\rightarrow\bar{\alpha}$) we realize that the $y-$component keeps invariant while the $x-$component $\left[\vec{\mathcal{J}}_{\mathcal{M}}\right]_x$ goes to $-\left[\vec{\mathcal{J}}_{\mathcal{M}}\right]_x$. In Figures \ref{F3} and \ref{F4} we can see plots of the $x$ and $y$ components of the current density  for the monolayer graphene coherent state, respectively.

\begin{figure}[ht] 
\begin{center}
\subfigure[]{\includegraphics[width=8cm, height=5.7cm]{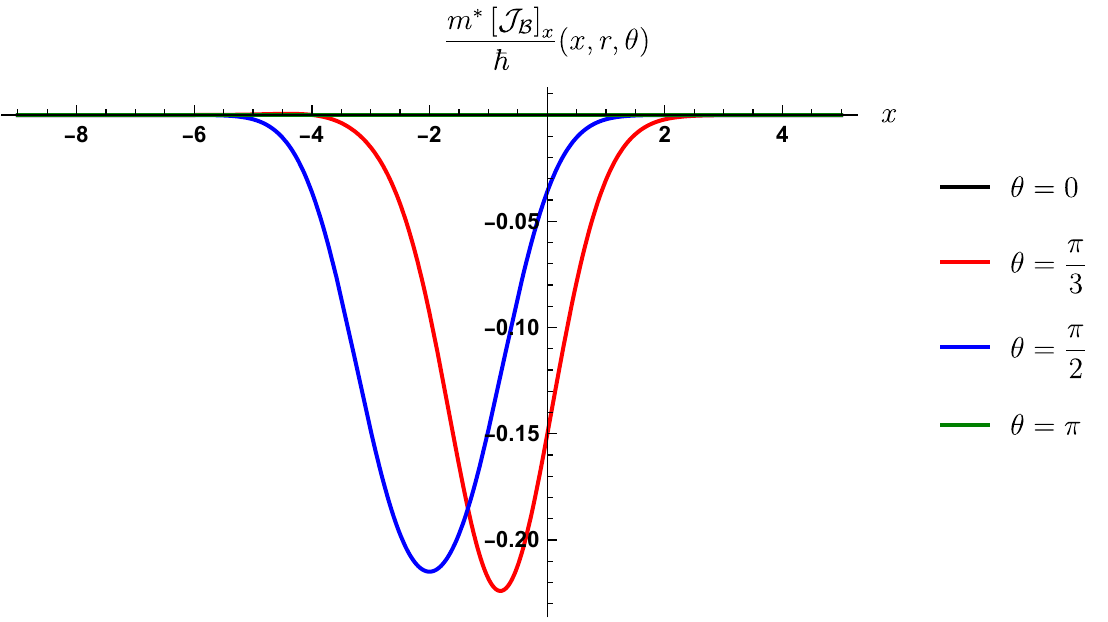}}
\subfigure[]{\includegraphics[width=8cm, height=5.7cm]{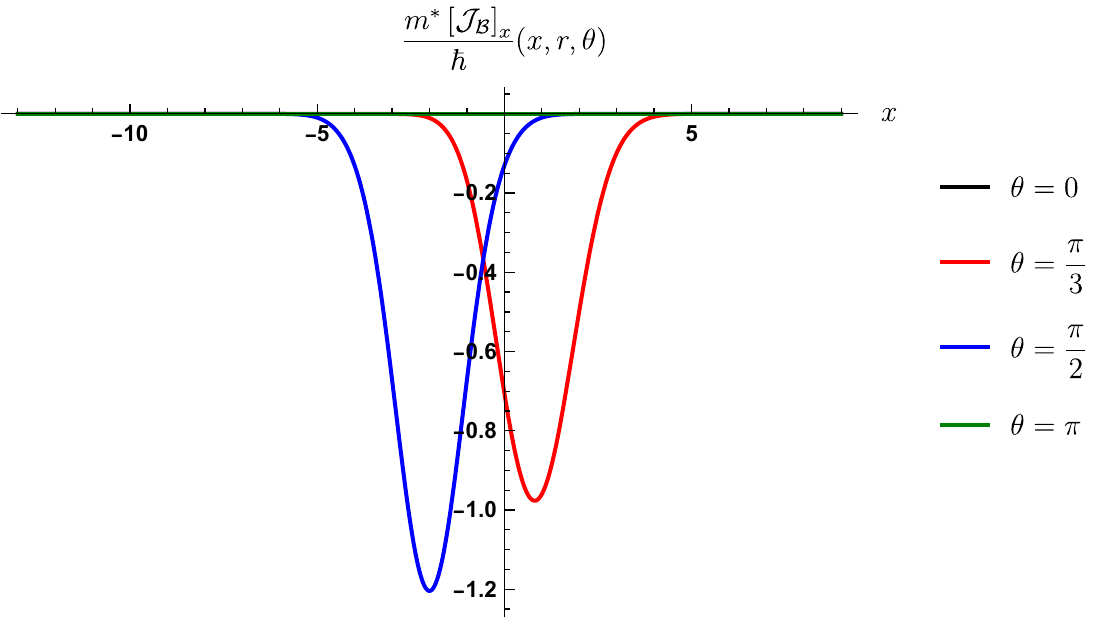}}
\subfigure[]{\includegraphics[width=8cm, height=5.7cm]{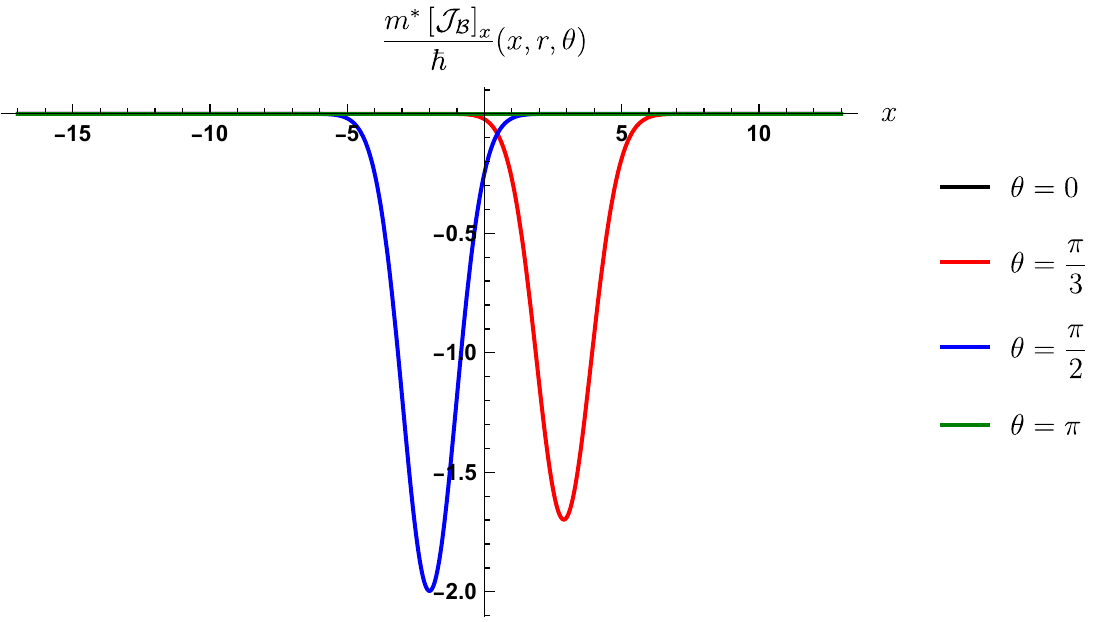}}
\caption{Plots of the $x-$component current density $\left[\mathcal{J}_\mathcal{B}\right]_x$ for the bilayer graphene coherent states (\ref{5.1.1}) as function of $x$ and $\theta$ for fixed values of $\omega=k=1$ and different values of $r$: (a) $r=1$; (b) $r=3$; (c) $r=5$.}
\label{F5}
\end{center}
\end{figure}

\begin{figure}[ht] 
\begin{center}
\subfigure[]{\includegraphics[width=8cm, height=5.7cm]{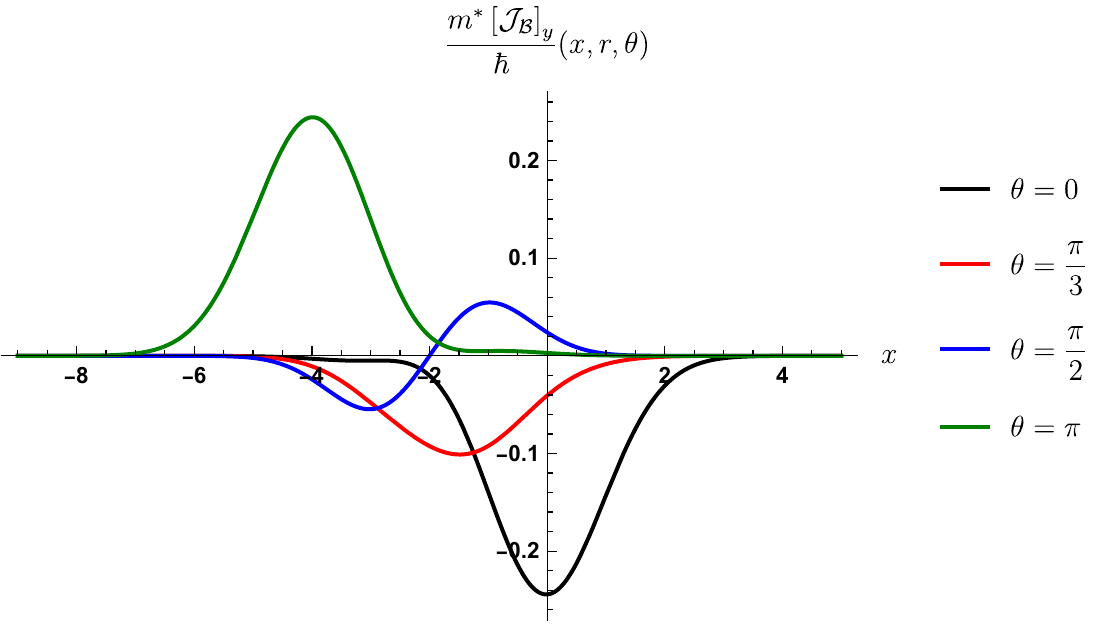}}
\subfigure[]{\includegraphics[width=8cm, height=5.7cm]{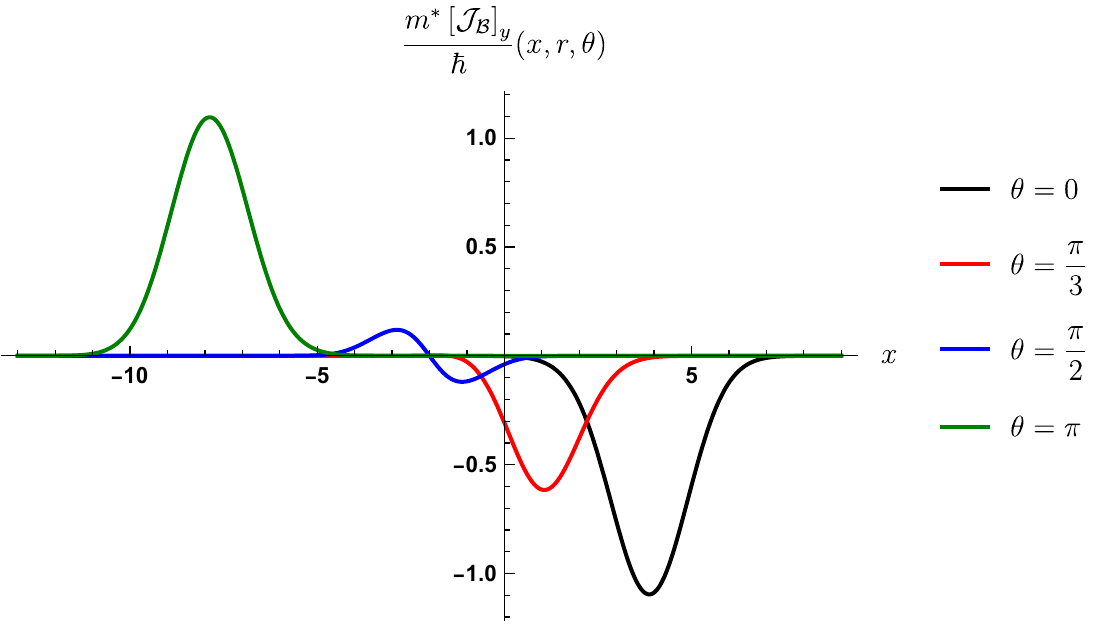}}
\subfigure[]{\includegraphics[width=8cm, height=5.7cm]{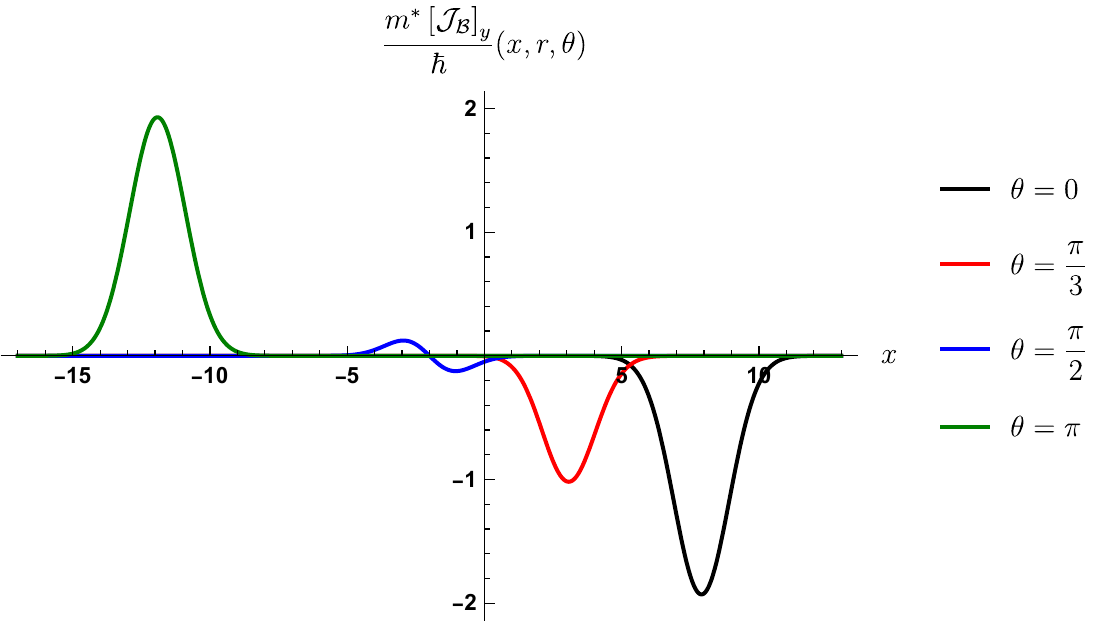}}
\caption{Plots of the  $y-$component current density $\left[\mathcal{J}_\mathcal{B}\right]_y$ for the bilayer graphene coherent states (\ref{5.1.1}) as function of $x$ and $\theta$ for fixed values of $\omega=k=1$ and different values of $r$: (a) $r=1$; (b) $r=3$; (c) $r=5$.}
\label{F6}
\end{center}
\end{figure}

For bilayer graphene, after some algebra we can see that the operator components (\ref{2.2.16}) turn out to be
\begin{equation}
j_x=
\begin{pmatrix} 
0 &\mathcal{L}_1^{-}\\
-\mathcal{L}_1^{+} & 0
\end{pmatrix}=\sqrt{\omega}
\begin{pmatrix} 
0 &\theta^{-}\\
-\theta^{+} & 0
\end{pmatrix},
\nonumber
\end{equation}
\begin{equation}
j_y=-i
\begin{pmatrix} 
0 &\mathcal{L}_1^{-}\\
\mathcal{L}_1^{+} & 0
\end{pmatrix}=
-i\sqrt{\omega}
\begin{pmatrix} 
0 &\theta^{-}\\
\theta^{+} & 0
\end{pmatrix}.
\label{5.3.3}
\end{equation}
In this way we obtain the following current density components:
\begin{align}
\left[\vec{\mathcal{J}}_{\mathcal{B}}\right]_x&=\frac{\hbar}{m^*}
\sqrt{\omega}e^{-r^2}
\sum_{n,m=0}^{\infty}\frac{r^{n+m}\mbox{sin}\left[(n-m)\theta\right]}{\sqrt{n!m!}}\mathrm{j}^{-}_{n,m}(x),\nonumber\\
\left[\vec{\mathcal{J}}_{\mathcal{B}}\right]_y&=-\frac{\hbar}{m^*}\sqrt{\omega}
e^{-r^2}
\sum_{n,m=0}^{\infty}\frac{r^{n+m}\mbox{cos}\left[(n-m)\theta\right]}{\sqrt{n!m!}}\mathrm{j}^{+}_{n,m}(x),
\label{5.3.4}
\end{align}
where now
\begin{equation}
\mathrm{j}^{\pm}_{n,m}(x)=\frac{\left(1-\delta_{m,0}-\delta_{m,1}\right)\sqrt{n}\psi^{-}_{m-2}(x)\psi^{-}_{n-1}(x)\pm\left(1-\delta_{n,0}-\delta_{n,1}\right)\sqrt{n-1}\psi^{-}_{m}(x)\psi^{-}_{n-1}(x)}{\sqrt{2^{2-\delta_{n,0}-\delta_{n,1}-\delta_{m,0}-\delta_{m,1}}}}.
\label{5.3.5}
\end{equation}
As for monolayer graphene, if $\theta\rightarrow -\theta$ ($\alpha\rightarrow\bar{\alpha}$) the $y-$ component keeps invariant while the $x-$component $\left[\vec{\mathcal{J}}_{\mathcal{B}}\right]_x$ goes to $-\left[\vec{\mathcal{J}}_{\mathcal{B}}\right]_x$. In Figures \ref{F5} and \ref{F6} we can see plots of the $x$ and $y$ components of the current density for bilayer graphene coherent state, respectively.
\subsubsection{Mean energy value}
Up to here we have constructed coherent states for monolayer and bilayer graphene through three different definitions. We have determined as well a family for which these definitions are mutually equivalent, as for the harmonic oscillator. However, according to the specific nature of the solutions some quantities, as the mean energy value, are different for each case.

\begin{figure}[ht] 
\begin{center}
\subfigure[]{\includegraphics[width=8cm, height=5.7cm]{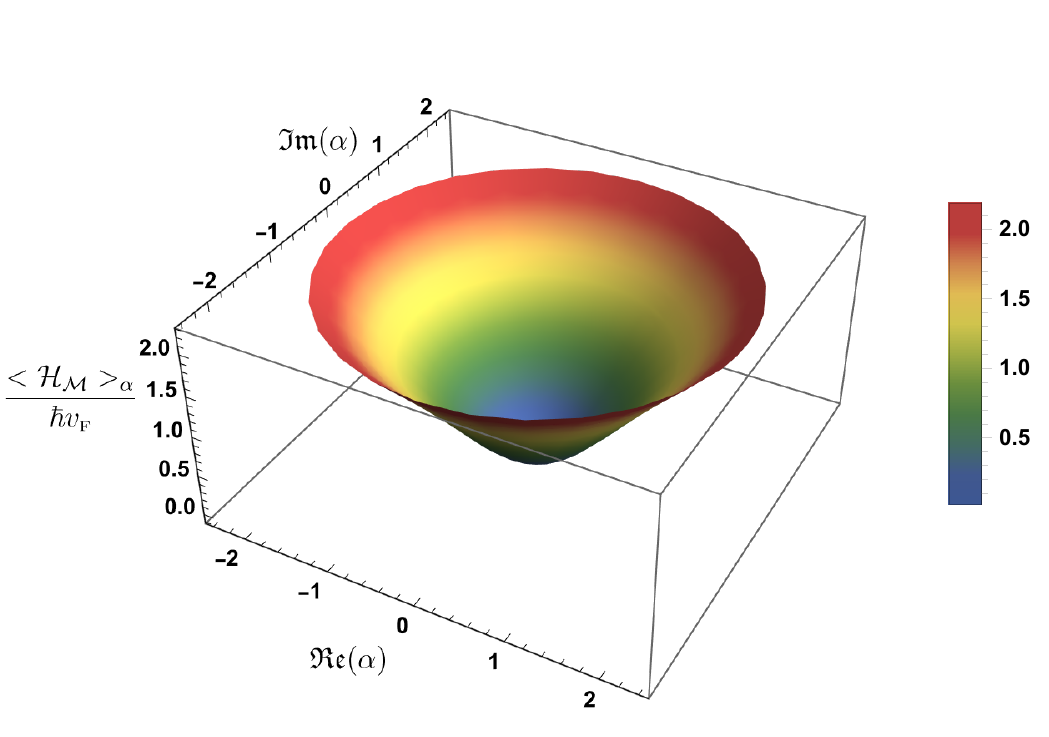}}
\subfigure[]{\includegraphics[width=8cm, height=5.7cm]{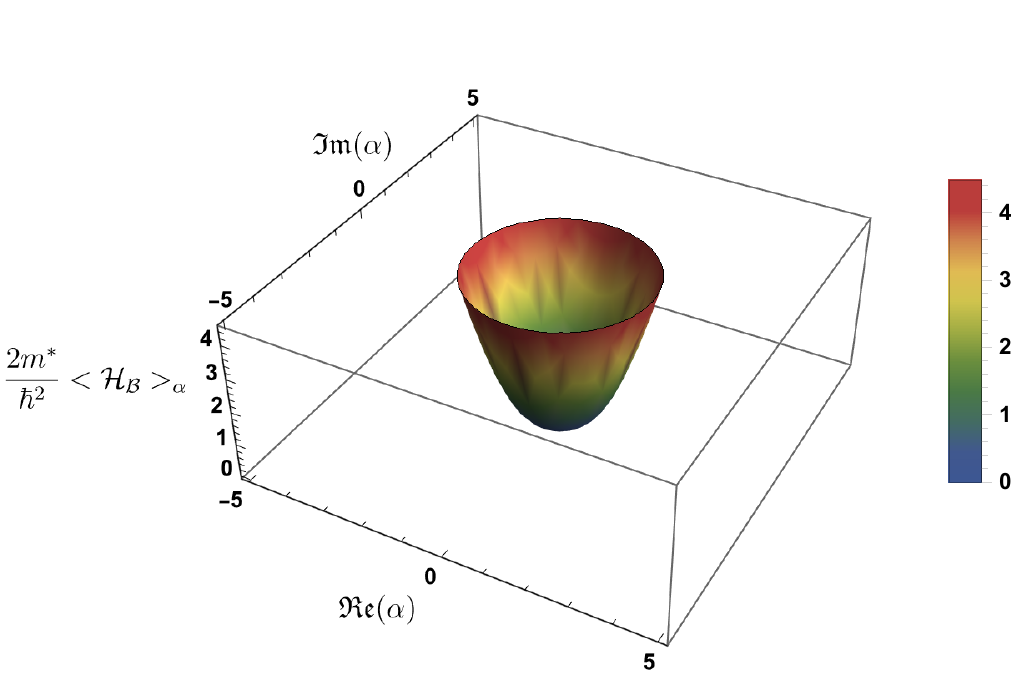}}
\caption{Mean energy value for the monolayer graphene coherent state (a) and the bilayer graphene coherent state (b) in (\ref{5.1.1}). In both cases the parameters have been fixed as $\omega=k=1$; we can see the reflexion symmetry with respect to $\theta$.}
\label{F7}
\end{center}
\end{figure}

In particular, the mean energy value $\braket{\mathcal{H}_{\mathcal{G}}}_{\alpha}=\bra{\Psi_{\alpha}}\mathcal{H}_{\mathcal{G}}\ket{\Psi_{\alpha}}$ for the coherent state (\ref{5.1.1}) with the eigenstates and eigenenergies (\ref{5.2}) and (\ref{5.3}) lead to
\begin{align}
\braket{\mathcal{H}_{\mathcal{M}}}_{\alpha}&=\hbar
v_{{\scriptscriptstyle F}}\sqrt{\omega}e^{-r^2}r^2\sum_{n=0}^{\infty}\frac{r^{2n}}{\sqrt{n+1} n!},\nonumber\\
\braket{\mathcal{H}_{\mathcal{B}}}_{\alpha}&=\frac{\hbar^2\omega}{2m^*}e^{-r^2}r^4\sum_{n=0}^{\infty}\frac{r^{2n}}{\sqrt{(n+2)(n+1)} n!}.\label{5.4.1}
\end{align}
As we can see, in both cases the mean energy value does not depend on the phase but only on the norm $|\alpha|=r$ of the complex number $\alpha$. In Figure \ref{F7}(a) it is shown a plot of the mean energy value for the monolayer graphene coherent state while for the bilayer graphene coherent state it is shown in Figure \ref{F7}(b).
\subsubsection{Heisenberg uncertainty principle}
Although the quadratures saturate the uncertainty product $\Delta Q_\mathcal{G}\Delta P_\mathcal{G}$ for our coherent states, it is important to stress that $Q_\mathcal{G}$ and $P_\mathcal{G}$ do not represent the position and momentum of the electron. Thus, we will calculate next the Heisenberg uncertainty relationship for the canonical position and momentum operators.

\begin{figure}[ht] 
\begin{center}
\includegraphics[width=8cm, height=5.7cm]{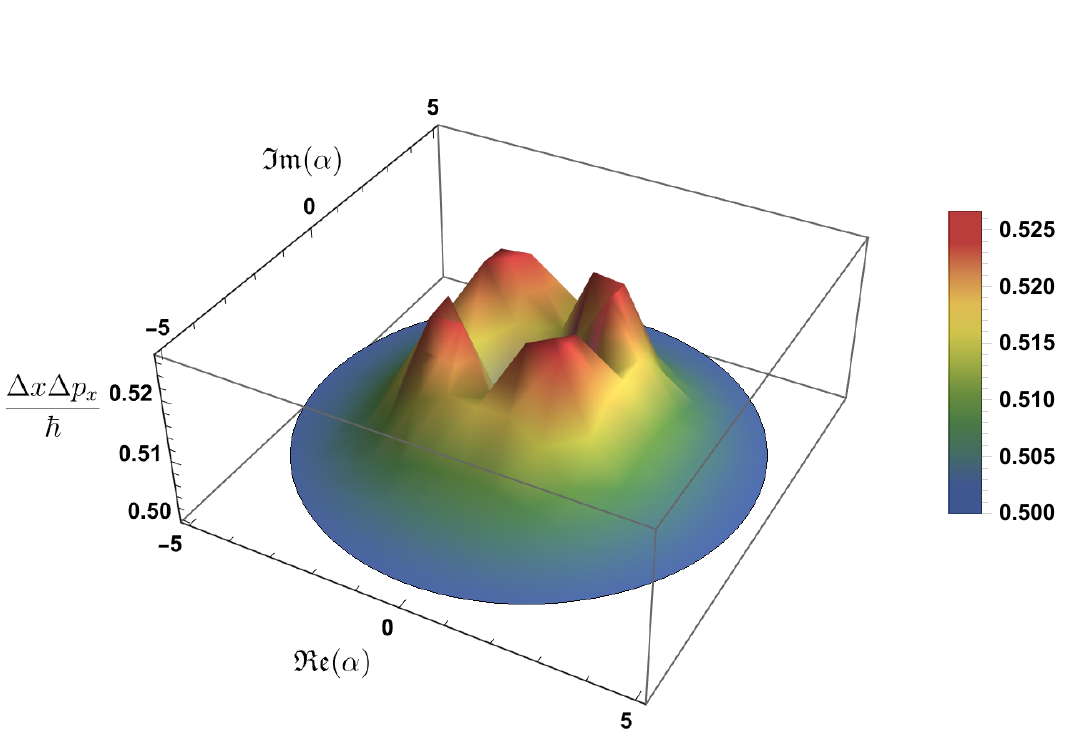}
\caption{Plot of the uncertainty product $\Delta x\Delta p_x$ as function of $\alpha$ for the monolayer graphene coherent sates (\ref{5.1.1}) in units of $\hbar$. It can be seen that for $r\rightarrow\infty$ this product is saturated (tends to $1/2$).} 
\label{F8}
\end{center}
\end{figure}

For the position $x$ and canonical momentum $p_x$  the Heisenberg uncertainty principle reads
\begin{equation}
\Delta x\Delta p_x=\hbar\Delta z\Delta p_z\geq \frac{\hbar}{2},
\label{5.5.1}
\end{equation}
where $z$ and $p_z$ are dimensionless position and momentum operators defined by 
\begin{align}
z=&\frac{\theta^++\theta^-}{\sqrt{2}},\nonumber\\
p_z=&i\frac{\theta^+-\theta^-}{\sqrt{2}}.
\label{5.5.2}
\end{align}
For the monolayer graphene coherent states (\ref{5.1.1}) it is straightforward to prove that the mean values of $z$ and $p_z$, as well as their squares, are given by
\begin{align}
\braket{z}_{\alpha}=&\frac{e^{-r^{ 2}}\mathfrak{Re}(\alpha)}{\sqrt{2}}\left(e^{r^{2}}+\sqrt{2}-1+\sum_{n=0}^{\infty}\frac{r^{2n+2}}{\sqrt{n!(n+2)!}}\right),\nonumber\\
\braket{z^2}_{\alpha}=&\frac{e^{-r^{2}}}{2}\left[1+2r^{2}e^{r^2}+\mathfrak{Re}(\alpha^2)\left(e^{r^2}+\sqrt{2}-1+\sum_{n=0}^{\infty}\frac{\sqrt{n+2}r^{2n+2}}{\sqrt{n!(n+3)!}}\right)\right],\nonumber\\
\braket{p_z}_{\alpha}=&\frac{e^{-r^{ 2}}\mathfrak{Im}(\alpha)}{\sqrt{2}}\left(e^{r^{2}}+\sqrt{2}-1+\sum_{n=0}^{\infty}\frac{r^{2n+2}}{\sqrt{n!(n+2)!}}\right),\nonumber\\
\braket{p_z^2}_{\alpha}=&\frac{e^{-r^{2}}}{2}\left[1+2r^{2}e^{r^2}-\mathfrak{Re}(\alpha^2)\left(e^{r^2}+\sqrt{2}-1+\sum_{n=0}^{\infty}\frac{\sqrt{n+2}r^{2n+2}}{\sqrt{n!(n+3)!}}\right)\right].
\label{5.5.3}
\end{align}
In the limit $\alpha\rightarrow0$ it turns out that $\Delta z=\Delta p_z=1/\sqrt{2}$ and thus in units of $\hbar$ $\Delta x\Delta p_x=1/2$ (see Figure \ref{F8}). 

\begin{figure}[ht] 
\begin{center}
\includegraphics[width=8cm, height=5.7cm]{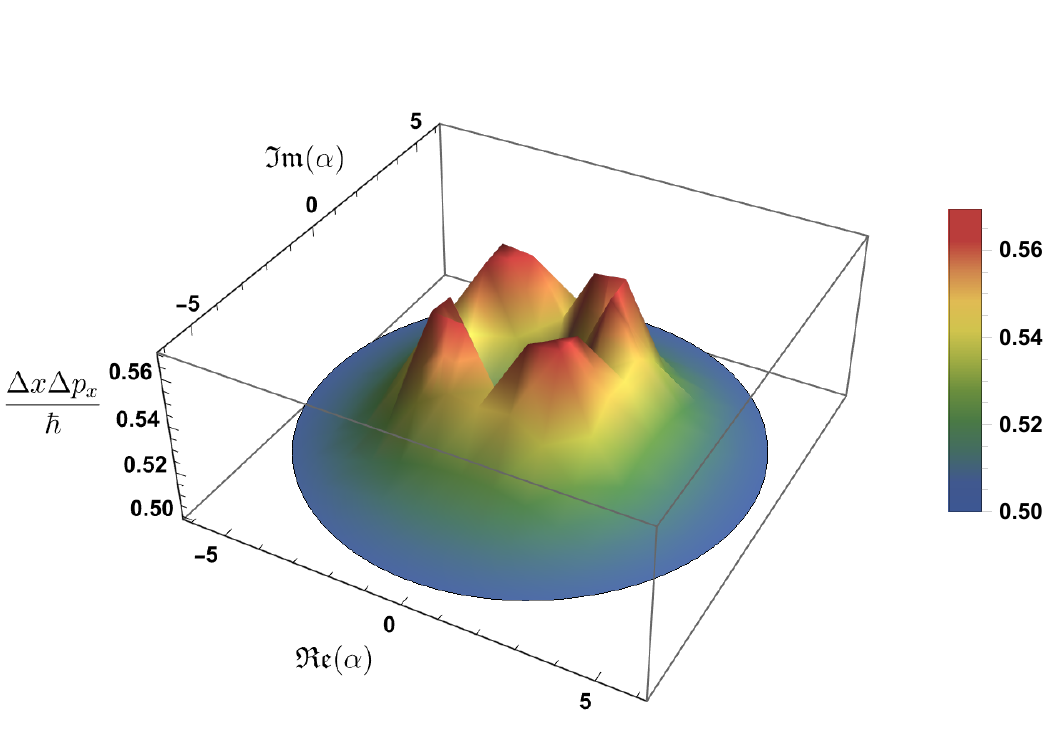}
\caption{Plot of the uncertainty product $\Delta x\Delta p_x$ as function of $\alpha$ for the bilayer graphene coherent sates (\ref{5.1.1}) in units of $\hbar$. It can be seen that for $r\rightarrow\infty$ this product is saturated (tends to $1/2$).} 
\label{F9}
\end{center}
\end{figure}

Now, for the bilayer graphene coherent states (\ref{5.1.1}) the corresponding mean values turn out to be
\begin{align}
\braket{z}_{\alpha}=&\frac{e^{-r^{ 2}}\mathfrak{Re}(\alpha)}{\sqrt{2}}\left[e^{r^{2}}+1+\left(\sqrt{2}-1\right)r^{2}+\sum_{n=0}^{\infty}\frac{\sqrt{n+1}r^{2n+4}}{\sqrt{(n+2)!(n+3)!}}\right],\nonumber\\
\braket{z^2}_{\alpha}=&\frac{e^{-r^{2}}}{2}\left[2(r^2+1)+(2r^2-1)e^{r^2}+\mathfrak{Re}(\alpha^2)\left(e^{r^2}+(\sqrt{2}-1)(1+r^2)+\sum_{n=0}^{\infty}\frac{r^{2n+4}}{\sqrt{n!(n+4)!}}\right)\right],\nonumber\\
\braket{p_z}_{\alpha}=&\frac{e^{-r^{ 2}}\mathfrak{Im}(\alpha)}{\sqrt{2}}\left[e^{r^{2}}+1+\left(\sqrt{2}-1\right)r^{2}+\sum_{n=0}^{\infty}\frac{\sqrt{n+1}r^{2n+4}}{\sqrt{(n+2)!(n+3)!}}\right],\nonumber\\
\braket{p_z^2}_{\alpha}=&\frac{e^{-r^{2}}}{2}\left[2(r^2+1)+(2r^2-1)e^{r^2}-\mathfrak{Re}(\alpha^2)\left(e^{r^2}+(\sqrt{2}-1)(1+r^2)+\sum_{n=0}^{\infty}\frac{r^{2n+4}}{\sqrt{n!(n+4)!}}\right)\right].
\label{5.5.4}
\end{align}
Once again, if $\alpha\rightarrow0$ then $\Delta z=\Delta p_z=1/\sqrt{2}$ and thus in units of $\hbar$ $\Delta x\Delta p_x=1/2$ (see Figure \ref{F9}). 
\subsubsection{Quantum fidelity}
It is well known that if the evolution operator $U(t,t_0)=e^{-i\mathcal{H}_{\mathcal{G}}(t-t_0)/\hbar}$ is applied to the coherent states (\ref{5.1.1}), the non-equidistant nature of the energy levels prevents to obtain a time evolution similar to that of the harmonic oscillator. However, in both cases it is possible to analyze the time evolution through the quantum fidelity, specially for bilayer graphene where approximate oscillation periods can be found, as will be shown below (see also \cite{Fernandez2020}).

Let us start by considering the time evolution of the monolayer graphene coherent states (\ref{5.1.1}), for which we have
\begin{equation}
\Psi_{\alpha}(x,y,t;\mathcal{M})=e^{-\frac{1}{2}r^2}\sum_{n=0}^{\infty}\frac{\alpha^{n}}{\sqrt{n}!}e^{-i\sqrt{n}t_1}
\Psi_n(x,y;\mathcal{M}),
\label{5.1.5.1}
\end{equation}
where $t_1=v_{{\scriptscriptstyle F}}\sqrt{\omega}t$ is a dimensionless time. On the other hand, for bilayer graphene it turns out that 
\begin{equation}
\Psi_{\alpha}(x,y,t;\mathcal{B})=e^{-\frac{1}{2}r^2}\sum_{n=0}^{\infty}\frac{\alpha^{n}}{\sqrt{n}!}e^{-i\sqrt{n(n-1)}t_2}
\Psi_n(x,y;\mathcal{B}),
\label{5.1.5.2}
\end{equation}
with $t_2=\hbar\omega t/2m^*$ being another dimensionless time; in both cases we have taken $t_0=0$. Additionally, we know that the quantum fidelity gives us a measure of the closeness between two states, which is defined by
\begin{equation}
F(\phi,\zeta)\equiv|\braket{\phi|\zeta}|^2.
\label{5.1.5.3}
\end{equation}
Thus, if $F(\phi,\zeta)=1$ it can be said that $\ket{\phi},\ket{\zeta}$ differ at most in a global phase factor and both represent the same quantum state. Keeping this in mind, one can say that if for some time $\tau$ the initial and the evolved state coincide, thus $\tau$ can be called a quasiperiod.

\begin{figure}[ht] 
\begin{center}
\subfigure[]{\includegraphics[width=8cm, height=5.7cm]{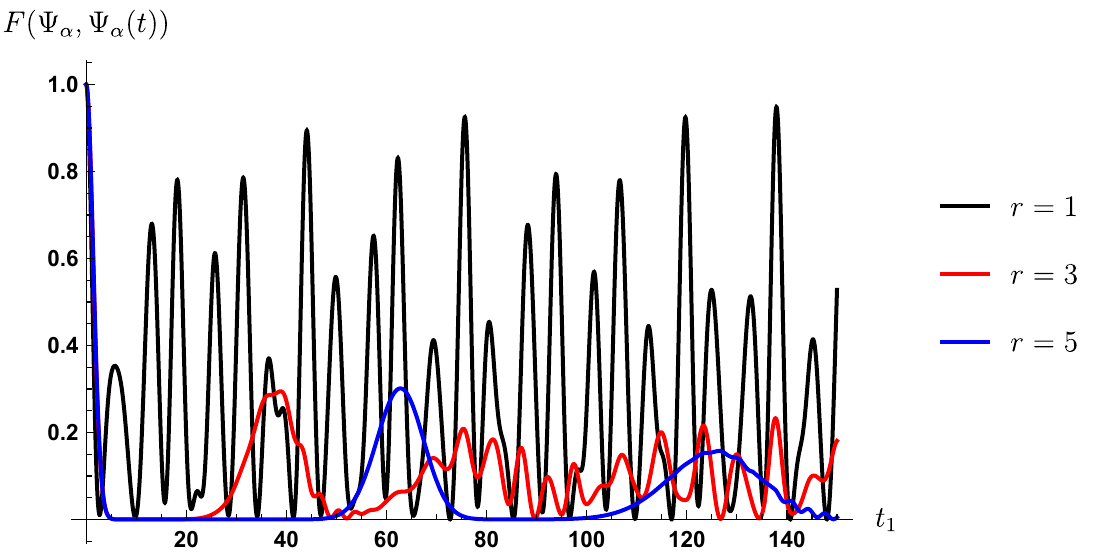}}
\subfigure[]{\includegraphics[width=8cm, height=5.7cm]{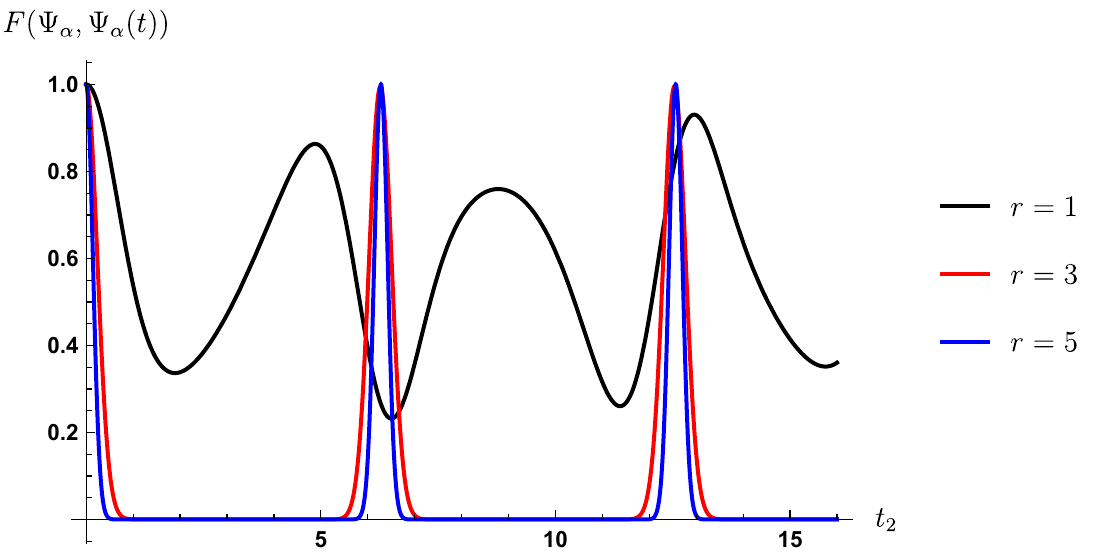}}
\caption{(a) Plots of the fidelity between the initial and the evolved state (\ref{5.1.1}) and (\ref{5.1.5.1}) for the monolayer graphene coherent states as function of the dimensionless time $t_1$. (b) Plots of fidelity between the initial and the evolved state (\ref{5.1.1}) and (\ref{5.1.5.2}) for the bilayer graphene coherent states as function of the dimensionless time $t_2$. In both cases the quantum fidelity becomes independent on the phase $\theta$.}
\label{F10}
\end{center}
\end{figure}

Now, for both coherent states (\ref{5.1.5.1}, \ref{5.1.5.2}) the fidelity is given by
\begin{equation}
F(\Psi_{\alpha},\Psi_{\alpha}(t))=e^{-2r^2}\sum_{n,m=0}^{\infty}\frac{r^{2(n+m)}}{n!m!}\mbox{cos}
\left[\left(h(n)-h(m)\right)t_{\mathcal{G}}\right],
\label{5.1.5.4}
\end{equation}
where $t_{\mathcal{G}}=\left\lbrace t_{\mathcal{M}}\equiv t_1,t_{\mathcal{B}}\equiv t_2,\right\rbrace$ and 
\begin{equation}
h(n)=
\left\lbrace
\begin{array}{c}
\sqrt{n}\quad\mbox{for}\quad \mbox{monolayer}\quad \mbox{graphene},\\
\\
\sqrt{n(n-1)}\quad\mbox{for}\quad \mbox{bilayer}\quad \mbox{graphene}.
\end{array}
\right.
\label{5.1.5.5}
\end{equation}
In Figure \ref{F10} we can see the quantum fidelity (\ref{5.1.5.4}) between the initial and the evolved state at time $t$ for the monolayer and bilayer graphene coherent states. In both cases such a quantity depends only on $r$ and the time ($t_1$ or $t_2$), and we can easily identify the values of $t_1$, $t_2$ that make $F(\Psi_{\alpha},\Psi_{\alpha}(t))$ going to $1$, which can be considered as the quasiperiods $\tau$.
\newpage
\textit{Discussion}

For bilayer graphene the eigenenergies can be written approximately as follows (see a similar approximation in \cite{PhysRevA.50.1814})
\begin{equation}
E_n=\frac{\hbar^2\omega}{2m^*}\left\lbrace
\begin{array}{c}
\quad 0\quad\mbox{for}\quad n=0,1,\\
\\
n-\frac{1}{2}+\mathcal{O}(\frac{1}{n})\quad\mbox{for}\quad n\geq2.
\end{array}
\right.
\label{5.1.5.6}
\end{equation}
Departing from a certain value $N$ the approximate expression (\ref{5.1.5.6}) can be supposed to be linear, thus we will have that
\begin{equation}
\Psi_{\alpha}(x,y,t)=\Psi^{'}_{\alpha}(x,y,t)+\gamma_{\alpha}(x,y,t,N),
\label{5.1.5.7}
\end{equation}
where
\begin{align}
\Psi^{'}_{\alpha}(x,y,t)&=e^{it_2/2}\left(e^{-\frac{r^2}{2}}\sum_{n=0}^\infty \frac{\alpha(t)^n}{\sqrt{n!}}\Psi_n(x,y)\right),\nonumber\\
\gamma_{\alpha}(x,y,t;N)&=e^{-\frac{r^2}{2}}\sum_{n=0}^{N-1} \frac{\alpha^n}{\sqrt{n!}}\left(e^{-i\sqrt{n(n-1)}t_2}-e^{-i(n-1/2)t_2}\right)\Psi_n(x,y),
\label{5.1.5.8}
\end{align}
and $\alpha(t)=\alpha e^{-it_2}$ (remember that $t_2$ is proportional to $t$). The previous expression allows us to write the fidelity as follows:
\begin{align}
F(\Psi_{\alpha},\Psi_{\alpha}(t))&=F(\Psi_{\alpha},\Psi^{'}_{\alpha}(t))+F(\Psi_{\alpha},\gamma_{\alpha}(t;N))+2
\mathfrak{Re}\left(\braket{\Psi^{'}_{\alpha}(t)|P_{\alpha}|\gamma_{\alpha}(t;N)}\right)\nonumber\\
&=e^{-4r^2\mbox{sin}^2(\frac{t_2}{2})}+F(\Psi_{\alpha},\gamma_{\alpha}(t;N))+2\mathfrak{Re}\left(\braket{\Psi^{'}_{\alpha}(t)|P_{\alpha}|\gamma_{\alpha}(t;N)}\right),
\label{5.1.5.9}
\end{align}
with $P_{\alpha}$ being the projector onto the subspace generated by the initial coherent state. In order to find the quasiperiods, the fidelity must be made approximately equal to 1, thus a value of $N$ is fixed and the equation is solved for $t_2$. Note that for a given $N$, regardless its value, if $r\rightarrow\infty$ then $\gamma_{\alpha}\rightarrow 0$. This means that $F(\Psi_{\alpha},\Psi_{\alpha}(t))=e^{-4r^2\mbox{sin}^2(\frac{t_2}{2})}$ in such a limit, which implies that the quasiperiod will be $\tau=2\pi$ 
and its integer multiples; we can see that this approximation agrees with the behavior observed in Figure \ref{10}(b) for $r>1$. Plots of the corresponding probability density at different times are shown in Figure \ref{11}.
\begin{figure}[ht] 
\begin{center}
\subfigure[]{\includegraphics[width=8cm, height=5.7cm]{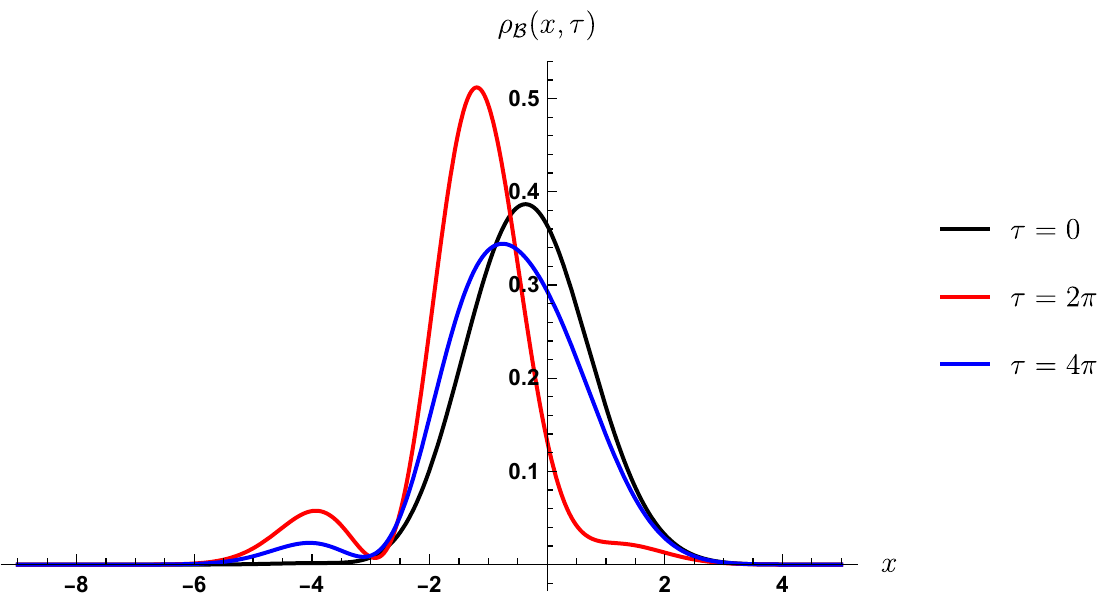}}
\subfigure[]{\includegraphics[width=8cm, height=5.7cm]{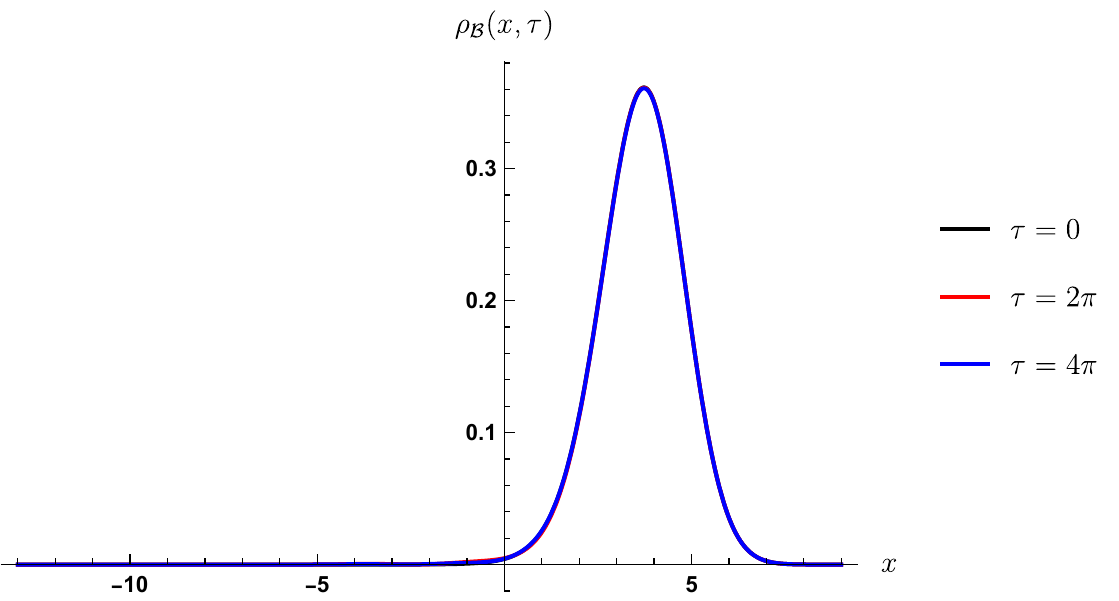}}
\subfigure[]{\includegraphics[width=8cm, height=5.7cm]{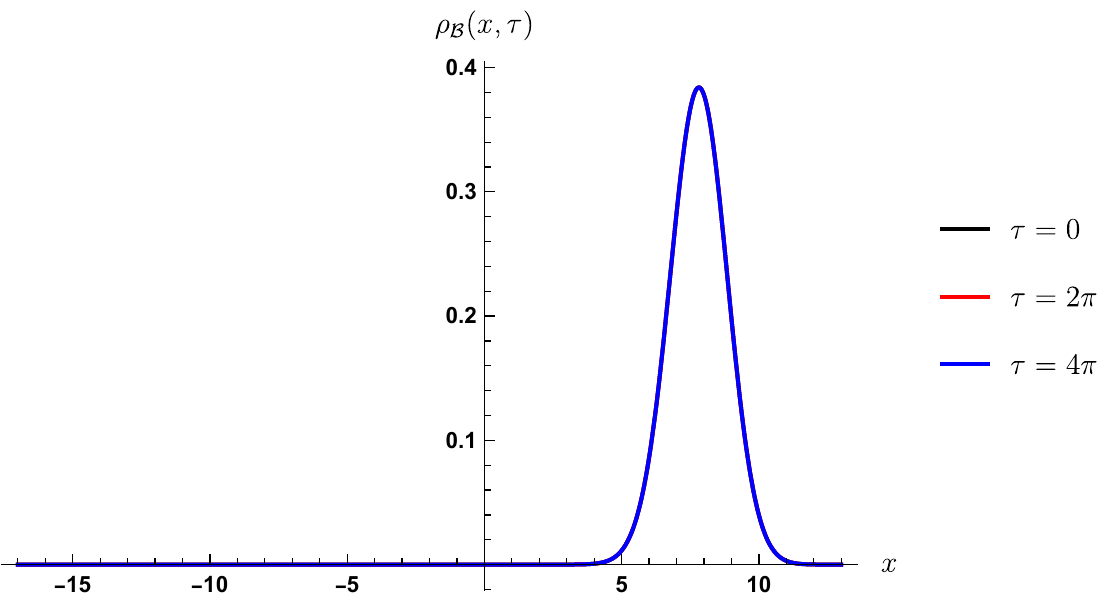}}
\caption{Plots of the probability densities for bilayer graphene coherent states at different multiples of the quasiperiod $\tau=2\pi$ and different values of $r$: (a) $r=1$; (b) $r=3$; (c) $r=5$. We can see that the different curves tend to be the same as $r$ increases. We have fixed $\theta=0$ and $\omega=k=1$.}
\label{F11}
\end{center}
\end{figure}
\section{Conclusions}
Graphene coherent states have been generated in several recent works \cite{Diaz-Bautista2017,D_az_Bautista_2019,Fernandez2020,Diaz-Bautista2021,CASTILLOCELEITA2020168287}, usually taking either the Barut-Girardello or the Gilmore-Perelomov definitions. In each of these works, annihilation operators with different properties have been used. This motivated us to analyze a different approach to construct them. We have started by finding out a family of creation and annihilation operators for the eigenstates associated with electrons in monolayer or bilayer graphene in an external magnetic field, in a similar way as it was done in \cite{Diaz-Bautista2017,Fernandez2020}. This allowed us to construct well-behaved ladder operators, such that they connect appropriately the eigenfunctions of the system Hamiltonian.

By making use of these ladder operators, we have derived and analyzed the coherent states through the Barut-Girardello, Gilmore-Perelomov and minimum uncertainty definitions. In the particular case of Gilmore-Perelomov coherent states, we have found suitable conditions for the commutator of the ladder operators to fulfill the Heisenber-Weyl algebra. This is an important result of this paper, since it allows us to employ the Baker-Hausdorff formula and to get a unitary displacement operator, in contrast with the operator obtained in \cite{Diaz-Bautista2021}.

After doing this, we have found a family of coherent states associated with the function choice $f(n)=1$, which allowed us to get the mutual equivalence between the different coherent states definitions. This is, in fact, the main difference with the previous works on graphene coherent states.

We have taken advantage of this equivalence and have determined some physical quantities, as the probability and current densities, time-evolution and quantum fidelity. Finally, in order to get an approximate oscillation period, as for the harmonic oscillator, we have carried out a fidelity analysis for the particular case of bilayer graphene coherent sates.
\bigskip

\noindent{\bf Acknowledgments.} This work was supported by CONACYT (Mexico), through the project FORDECYT-PRONACES/61533/2020. 
DOC especially thanks Conacyt for economic support through the PhD scholarship number 703048.
\bibliographystyle{Nature}
\bibliography{biblio}

\end{document}